\documentclass[journal]{IEEEtran}

\ifCLASSINFOpdf
\else
\fi

\usepackage{graphicx}
\usepackage{amsmath}
\usepackage{bm}
\usepackage{multirow}

\begin{document}
\title{Power-Constrained Secrecy Rate Maximization for Joint Relay and Jammer Selection Assisted Wireless Networks}

\markboth{IEEE Transactions on Communications (ACCEPTED TO APPEAR)}%
{H. Guo \MakeLowercase{\textit{et al.}}: Power-Constrained Secrecy Rate Maximization for Joint Relay and Jammer Selection Assisted Wireless Networks}

\author{Haiyan~Guo,
        Zhen~Yang,
        Linghua~Zhang,
        Jia~Zhu,
        and~Yulong~Zou,~\IEEEmembership{Senior~Member,~IEEE}% <-this % stops a space

\thanks{Manuscript received July 27, 2016; revised November 07, 2016; accepted December 31, 2016. The editor coordinating the review of this paper and approving it for publication was Prof. J. Yuan.}
\thanks{This work was partially supported by the National Natural Science Foundation of China (Grant Nos. 61302152, 61302104, 61401223, 61522109 and 61671252), the Major Science Research Project of Jiangsu Provincial Education Department (Grant Nos. 13KJA510003, 14KJA510003 and 15KJA510003), the Natural Science Foundation of Jiangsu Province (Grant Nos. BK20140887 and BK20150040), the Priority Academic Program Development of Jiangsu Higher Education Institutions (PAPD) and Research Foundation for Advanced Talents, Nanjing University of Posts and Telecommunications (No. NY215020)}% <-this % stops a space
\thanks{H. Guo and L. Zhang are with the Key Laboratory of Broadband Wireless Communication and Sensor Network Technology (Ministry of Education) and the College of Internet of Things, Nanjing University of Posts and Telecommunications, Nanjing 210003, China (e-mail: \{guohy, zhanglh\}@njupt.edu.cn).}% <-this % stops a space
\thanks{Z. Yang is with the Key Laboratory of Broadband Wireless Communication and Sensor Network Technology (Ministry of Education), Nanjing University of Posts and Telecommunications, Nanjing 210003, China (e-mail: yangz@njupt.edu.cn).}% <-this % stops a space
\thanks{Y. Zou and J. Zhu are with the Key Laboratory of Broadband Wireless Communication and Sensor Network Technology (Ministry of Education) and the School of Telecommunications and Information Engineering, Nanjing University of Posts and Telecommunications, Nanjing 210003, China (e-mail: \{yulong.zou, jiazhu\}@njupt.edu.cn).}% <-this % stops a space
\thanks{Corresponding author is Yulong Zou.}% <-this % stops a space
}

\maketitle

\begin{abstract}
In this paper, we examine the physical layer security for cooperative wireless networks with multiple intermediate nodes, where the decode-and-forward (DF) protocol is considered. We propose a new joint relay and jammer selection (JRJS) scheme for protecting wireless communications against eavesdropping, where an intermediate node is selected as the relay for the sake of forwarding the source signal to the destination and meanwhile, the remaining intermediate nodes are employed to act as friendly jammers which broadcast the artificial noise for disturbing the eavesdropper. We further investigate the
power allocation among the source, relay and friendly jammers for maximizing the secrecy rate of proposed JRJS scheme and derive a closed-form sub-optimal solution. Specificially, all the intermediate nodes which successfully decode the source signal are considered as relay candidates. For each candidate, we derive the sub-optimal closed-form power allocation solution and obtain the secrecy rate result of the corresponding JRJS scheme. Then, the candidate which is capable of achieving the highest secrecy rate is selected as the relay. Two assumptions about the channel state information (CSI), namely the full CSI (FCSI) and partial CSI (PCSI), are considered. Simulation results show that the proposed JRJS scheme outperforms the conventional pure relay selection, pure jamming and GSVD based beamforming schemes in terms of secrecy rate. Additionally, the proposed FCSI based power allocation (FCSI-PA) and PCSI based power allocation (PCSI-PA) schemes both achieve higher secrecy rates than the equal power allocation (EPA) scheme.
\end{abstract}


% \fi
\IEEEpeerreviewmaketitle

\section{Introduction}
\IEEEPARstart{D}{ue} to the broadcast nature of wireless medium, the confidentiality of wireless communications is vulnerable to an eavesdropping attack [1]. Physical layer security is emerging as a promising paradigm against eavesdropping and has attracted considerable attention recently. Different from the conventional encryption methods [2], it mainly exploits the physical characteristics of wireless channels to enhance the security of the signal transmission from its source to the intended destination. Physical layer security work was pioneered by Wyner for a discrete memoryless wiretap channel [3], which later on, was extended to a Gaussian wiretap channel in [4]. The secrecy capacity which is the maximum of secrecy rate  was developed to evaluate the physical-layer security performance. In [4], it showed that the  secrecy capacity can be expressed as the difference between the channel capacity from source to destination (referred to as main channel) and the channel capacity from source to eavesdropper (called wiretap channel) in the presence of an eavesdropper.\\
\indent Multiple-input multiple-output (MIMO) [5]-[11] has been widely considered as an effective way to improve physical layer security by exploiting multiple antennas at both the transmitter and the receiver. However, due to the battery power and terminal size limitations, it may be difficult to implement multiple antennas in some cases e.g. handheld terminals and wireless sensors. As a consequence, node cooperation is considered to be an effective means of enhancing the secrecy capacity by allowing the network nodes to share each other's antennas [12]-[31]. Generally, there are two efficient ways to make use of multiple intermediate nodes: cooperative beamforming and cooperative jamming. To be specific, cooperative bearmforming helps to improve the capacity of main channel, which may include relay selection and beamforming [12]-[18]. By contrast, cooperative jamming aims to degrade the capacity of the wiretap channel by sending artificial noise to interfere with the eavesdropper without affecting the legitimate destination [19]-[21]. Both  cooperative beamforming and jamming are capable of increasing the secrecy capacity of wireless communications in the presence of an eavesdropper.\\
\indent In [12] and [13], the authors studied the cooperative beamforming and cooperative jamming separately, where all the intermediate nodes are used either to assist the source transmission or to send a jamming signal for interfering with the eavesdropper. The transmit power at the source and the weights of intermediate nodes are determined to maximize the secrecy rate subject to a total power constraint. In [14], an optimal relay selection scheme was proposed by taking into account the channel state information (CSI) of both the main channel and wiretap channel, which is shown to achieve a better security performance than the conventional relay selection and combining methods. Almost meanwhile, cooperative jamming was also studied for increasing the secrecy rate. In [19], optimal cooperative jamming was studied with an emphasis on how to guarantee a positive secrecy rate. In [20], the use of a jammer was proposed to send artificial noise along with confidential signal for confusing the wiretap link and enhancing the main link simultaneously.\\
\indent The aforementioned studies are based on the ideal assumption that the instantaneous CSI of both main and wiretap links are known. Considering that obtaining the instantaneous CSI of the wiretap channel is challenging in many cases, Y. Su et al. [22] investigated the relay selection and power allocation by assuming that the correlation between the real and outdated CSIs is known without requiring the instantaneous CSI. In [23], X. Gong et al. presented  a robust relay beamforming scheme which maximizes the worst-case transmit power of the artificial noise with imperfect CSI. Moreover, in [24], F. S. Al-Qahtani et al. studied the impact of CSI feedback delay on the secrecy performance of three different diversity combining methods under the imperfect CSI case. In addition, C. Wang et al. [25] investigated the relay selection and power allocation between two sources and mulitiple relays for a two-way amplify-and-forward (AF) wireless network based on statistical channel information.\\
\indent In order to take both advantages of cooperative beamforming and cooperative jamming, various joint relay selection and jamming (JRJS) schemes were proposed for further enhancing the wireless secrecy performance, where some intermediate nodes help the legitimate transmission while some other nodes may jam the eavesdropper [26]-[31]. More specifically, the authors of [26] and [27] selected one intermediate node as the relay and two other nodes as jammers in a two-way cooperative network, achieving a better secrecy performance than conventional non-jamming schemes. Two different channel knowledge assumptions were considered in [27]: 1) an instantaneous CSI of the eavesdropping channel is assumed to be known, and 2) only an average CSI of the wiretap channel is known. By contrast, in [28] and [29], one intermediate node was chosen as the jammer to send artificial noise and the other intermediate nodes were considered to be relays for data transmission. A second-order convex cone programming algorithm was proposed in [29] to obtain sub-optimal beamforming weights and power allocation without eavesdropper's CSI. In [30], an intermediate node that achieves the highest instantaneous signal-to-noise ratio (SNR) of the second hop was selected as the relay and a node of minimizing the interference imposed on the destination was used as the jammer, where an outdated estimated CSI is assumed. Additionally, an opportunistic relaying scheme with the artificial jamming was proposed in [31], where the power allocation between the confidential signal and jamming signals is optimized to maximize the ergodic secrecy rate. \\
\indent In this paper, we propose a new JRJS scheme to enhance the physical layer security. Specifically, an intermediate node which can successfully decode the source signal is selected to act as the relay for assisting the source transmission, while the remaining intermediate nodes are employed as friendly jammers to transmit null-steering artificial noise for confusing the eavesdropper without affecting the legitimate destination. We select one intermediate node as the relay to exploit the full spatial diversity and the others as jammers for the sake of reducing the complexity of JRJS without an extra jammer selection. In addition, if we select multiple intermediate nodes as the relays to help signal transmission in different channels, there would be more links where the eavesdropper can tap the legitimite transmission. Moreover, if the multiple relays share a common channel to forward their re-encoded outcomes, the corresponding symbol-level synchronization among the spatially distributed relays is very complex.\\
\indent Considering that the secrecy performance would be affected by the power allocation among the source, relay and jammers, we further formulate a power allocation optimization problem to maximize the secrecy rate of the proposed JRJS scheme subject to a total power constraint and derive a closed-form sub-optimal power allocation solution. It is noted that the total power constraint is widely used in literature e.g., [12]-[13], [15]-[17], [21], [25] and [30]-[31]. In our proposed JRJS scheme, the relay selection and power allocation are jointly optimized by taking into account both the transimission rate and secrecy rate.  Since the instantaneous CSI may be unavailable in some practical scenarios, we consider two different CSI cases in this paper, which are called the full CSI (FCSI) and partial CSI (PCSI), respectively. The FCSI assumes that the instantaneous CSI of both main channel and wiretap channel is known, while the PCSI only requires the main channel's instantaneous CSI along with the statistical wiretap channel knowledge. \\
\indent  The main contributions of our work are summarized as follows. \\
\indent Firstly, the relay and jammer selection as well as the power allocation are jointly carried out to maximize the secrecy rate of legitimate transmission. To be specific, for each relay candidate, we derive an optimal power allocation solution and obtain the corresponding secrecy rate result. After that, the candidate that maximizes the secrecy rate is selected as the relay. Secondly, considering that the secrecy performance would be affected by the power allocation among the source, relay and jammers, we formulate a power allocation problem for maximizing the secrecy rate of proposed JRJS scheme subject to a total power constraint. Moreover, we derive closed-form sub-optimal solutions to the formulated power allocation problem under FCSI and PCSI assumptions, respectively. \\
\indent The remainder of this paper is organized as follows. In Section II, the system model and secrecy rate analysis are presented. In Section III, we derive the power allocation schemes under the FCSI and PCSI assumptions, respectively. Numerical evaluation results are described in Section IV and main conclusions are drawn in Section V.\\

\section{System Model and Secrecy Rate Analysis}
\subsection{System Model}
We consider a decode-and-forward (DF) cooperative wireless network consisting of one source $S$, one destination $D$ and $M$ intermediate nodes in the presence of an eavesdropper $E$ as shown in Fig. 1. Each node in the whole network is only equipped with a single antenna. The solid and dash lines represent the main and wiretap links, respectively. Both the main and wiretap links are modeled as Rayeigh fading channels and the thermal noise received at any node is modeled as a complex Gaussian random variable with zero mean and variance $\sigma_{n}^2$ . We assume that there is a direct link from the source to the eavesdropper and no direct link from the source to the destination. $M$ intermediate nodes are exploited to assist the signal transmission from the source to the destination. \\
\indent In this paper, we propose a new JRJS scheme, where the intermediate nodes are divided into two groups: one relay node $R$ and $M-1$ jammers  $J_i,i=1,2,...,M-1$, as shown in Fig. 1. More specifically, in the first phase, the source node broadcasts a signal $s$  with power $P_s$  to $M$  intermediate nodes, which all attempt to decode their received signals. Due to the broadcasting nature, the signal $s$ is also received by the eavesdropper. For convenience, the intermediate nodes which succeed in decoding the source signal are denoted by a decoding set $\cal{D}$. In the second phase, we select one intermediate node from $\cal{D}$ as the relay $R$ to re-encode and transmit its decoded outcome to the destination with power $P_r$. Meanwhile, the remaining $M-1$ intermediate nodes are enabled to act as friendly jammers $J_i,i=1,2,...,M-1$, to transmit artificial noise $z_i$  for confusing the eavesdropper. We assume that the distances among intermediate nodes are much smaller than the distances between intermediate nodes and the source/destination/eavesdropper following [30]. This assumption is reasonable for both WSNs and MANETs associated with a clustered relay configuration. Under this assumption, the corresponding path losses among the relay and jammers are approximately the same, hence we simplify our work without considering path loss. \\
\indent In this paper, we assume that the whole wireless network is to operate under a total power constraint following [12]-[13], [15]-[17], [21], [25] and [30]-[31], which is widely adopted in both secrecy performance analysis and optimal design. For example, as indicated in [17], half of the maximum power budget is allocated to the two transceivers and the remaining half will be shared among all the relay nodes for symmetric relaying schemes. In [30], it pointed out the secrecy performance is effected by power allocation between the relay and jammers. Futhermore, from a network design of view, such an assumption allows to optimize total power consumed in the whole network. In addition, the solved transmit power at the relay also provides a guideline for the transmit power of individual relays. The optimal power allocation among the source, relay and jammers will be addressed in following Section III. \\
\begin{figure}[htb!]
\centering
\includegraphics[scale=0.4]{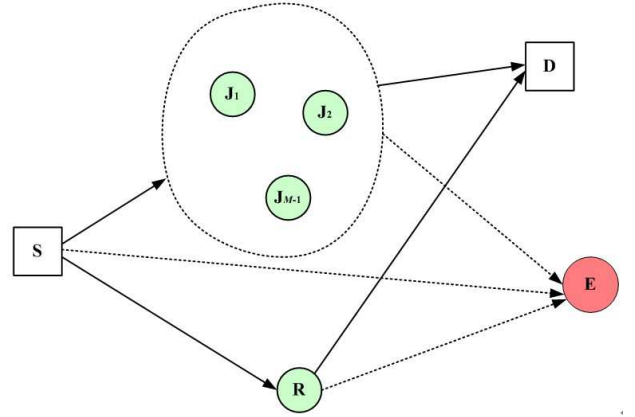}
\caption{A cooperative wireless network consisting of one source, one destination, one eavesdropper and $\textsl{M}$ intermediate nodes.}
\label{fig_1}
\end{figure}
\indent In the first phase, the source transmits a signal $s( E ( |s|^2 )=1)$ with power $P_s$. The intermediate node $i$ and eavesdropper $E$ thus receive
\begin{equation}\label{(1)}
y_i=\sqrt{P_s} h_{si} s+n_i
\end{equation}
\begin{equation}\label{(2)}
y_e^{(1)}=\sqrt{P_s} h_{se} s+n_e^{(1)}
\end{equation}
where $h_{si}$ and $h_{se}$ represent a fading coefficient of the channel from the source to the intermediate node $i$ and eavesdropper $E$, respectively, $n_i \in \mathcal{CN}(0,\delta_{n}^2)$ is additive white Gaussian noise (AWGN) at the intermediate node $i$, and $n_e^{(1)} \in \mathcal{CN}(0,\delta_{n}^2)$ is  AWGN at $E$.\\
\indent In the second phase, the relay $R$ is selected from the decoding set $\cal{D}$ for signal transmission with power $P_r$. Meanwhile, each jammer $J_i$  transmits an artificial noise signal $z_i$.  It needs to be pointed out that the vector of the artificial noise $\bm{z}=\lbrack z_1 \quad z_2 \quad ... \quad z_{\textsl{M}-1} \rbrack ^{\rm{T}}$ should be normalized. The total transmit power at all the jammers is denoted as $P_z$. Therefore, the received signal at the destination is expressed as
\begin{equation}\label{(3)}
y_d=\sqrt{P_r} h_{rd} s+\sum_{i=1,2,...,\textsl{M}-1}{\sqrt{P_z} h_{j_id}z_i}+n_d
\end{equation}
where $h_{rd}$ represents a fading coefficient of the channel from the relay $R$ to destination, $h_{j_id}$  represents a fading coefficient of the channel from the jammer $\textsl{J}_i$  to  destination and $n_d \in \mathcal{CN}(0,\delta_{n}^2)$ represents AWGN at the destination. The received signal at the eavesdropper is expressed as
\begin{equation}\label{(4)}
y_e^{(2)}=\sqrt{P_r} h_{re} s+\sum_{i=1,2,...,\textsl{M}-1}{\sqrt{P_z} h_{j_ie}z_i}+n_e^{(2)}
\end{equation}
where $h_{re}$ represents a fading coefficient of the channel from the relay to eavesdropper, $h_{j_ie}$  represents a fading coefficient of the channel from the jammer $J_i$  to eavesdropper and $n_e^{(2)} \in \mathcal{CN}(0,\delta_{n}^2)$ represents AWGN at the eavesdropper.\\
\indent For simplicity, let us define $\bm{h}_d=\lbrack h_{j_1d} \quad h_{j_2d} \quad ... \quad h_{j_{\textsl{M}-1}d} \rbrack ^{\rm{H}}$ and $\bm{h}_e=\lbrack h_{j_1e} \quad h_{j_2e} \quad ... \quad h_{j_{\textsl{M}-1}e} \rbrack ^{\rm{H}}$. Considering that the artificial noise should not interfere with $\textsl{D}$, we design $\bm{z}$  as a normalized signal onto the null space of $\bm{h}_d$ so that $\bm{h}_d^{\rm{H}}\bm{z}=\sum_{i=1,2,...,\textsl{M}-1}{h_{j_id}z_i}=0$. As a consequence, we obtain (3) and (4) as
\begin{equation}\label{(5)}
y_d=\sqrt{P_r} h_{rd} s+n_d
\end{equation}
and
\begin{equation}\label{(6)}
y_e^{(2)}=\sqrt{P_r} h_{re} s+\sqrt{P_z} \bm{h}_e^{\rm{H}}\bm{z}+n_e^{(2)}
\end{equation}
which completes the signal modeling of the JRJS scheme.

\subsection{Secrecy Rate Analysis}
\indent In [32], a variable transmission parameter scheme where the source and relay use different codebooks with different code rates is proposed for DF relay networks to further improve the secrecy performance. In [33], a framework is provided to determine the wiretap code rates to achieve the locally maximum effective secrecy throughput. In this paper, we still consider the case where the source and relay transmit signal using the same code with the same code rate $R_d$ in both main and wiretap links and focus on how to allocate transmit power among the source, relay and all the jammers to maximize the secrecy rate of the proposed JRJS scheme. Although we do not propose a new transmission parameter design scheme in this paper, we take  into account the effect of $R_d$ on the secrecy rate and set $R_d$ to different values based on the values of $\frac{P}{\delta_n^2}$ by experimental experience, which will be shown in the following experimental section. \\
\indent When the source and relay transmit the same code with $R_d$, the rate of DF transmission from source $S$  via the relay $R$ to destination $D$, $C_d$, is obtained as
\begin{equation}\label{(7)}
C_d=\mathrm{min}(C_{sr},C_{rd})
\end{equation}
where $C_{sr}$ and $C_{rd}$ represent the rate of the link from $S$  to $R$ and the rate of the link from $R$  to $D$, respectively, which are given as
\begin{equation}\label{(8)}
C_{sr}=\frac {1} {2} \log_2(1+\frac {|h_{sr}|^2 P_s} {\delta_n^2})
\end{equation}
and
\begin{equation}\label{(9)}
C_{rd}=\frac {1} {2} \log_2(1+\frac {|h_{rd}|^2 P_r} {\delta_n^2}).
\end{equation}
It is noted that the relay $R$ is selected from the decoding set $\cal{D}$. That is, it satisfies that
\begin{equation}\label{(10)}
\log_2(1+\frac {|h_{sr}|^2 P_s}{\delta_n^2}) \geq R_d .
\end{equation}
Combining (7)-(9), the rate of the link from $S$ to $D$ is given by
\begin{equation}\label{(11)}
C_{d}=\frac {1} {2} \log_2(1+\frac {\mathrm{min}(|h_{sr}|^2 P_s,|h_{rd}|^2 P_r)} {\delta_n^2}).
\end{equation}
\indent Combing (5) and (6), we obtain the rate of the link from $S$ to $E$ as
\begin{equation}\label{(12)}
C_{e}=\frac {1} {2}\log_2(1+\frac {|h_{se}|^2 P_s}{\delta_n^2}+\frac {|h_{re}|^2 P_r} {\delta_n^2+P_z \bm{h}_e^{\textrm{H}}\bm{z}\bm{z}^{\textrm{H}}\bm{h}_e}).
\end{equation}
For simplicity, let us define the instantaneous SNR at destination as a function of the relay $R$, $P_s$ and $P_r$
\begin{equation}\label{(13)}
\gamma_d(R,P_s,P_r)=\frac {\mathrm{min}(|h_{sr}|^2 P_s,|h_{rd}|^2 P_r)} {\delta_n^2}
\end{equation}
and the instantaneous SNR at eavesdropper as a function of $R$, $P_s$, $P_r$ and $P_z$
\begin{equation}\label{(14)}
\gamma_e(R,P_s,P_r,P_z)=\frac {|h_{se}|^2 P_s}{\delta_n^2}+\frac {|h_{re}|^2 P_r} {\delta_n^2+P_z \bm{h}_e^{\textrm{H}}\bm{z}\bm{z}^{\textrm{H}}\bm{h}_e}.
\end{equation}
Combining (11)-(14), the secrecy rate of the source-destination transmission relying on the proposed JRJS scheme is given by
\begin{equation}\label{(15)}
C_s=(C_d-C_e)^+=\frac {1} {2} \log_2 (\frac { 1+\gamma_d(R,P_s, P_r)} {1+\gamma_e(R,P_s,P_r,P_z)})^+
\end{equation}
where $(.)^+$ is a function such that $(x)^+=x$ if $x>0$ and $(x)^+=0$ if $x\leq0$.\\

\section{Power Allocation of Proposed JRJS Scheme}
\subsection{FCSI based Power Allocation (FCSI-PA) Scheme}
In this subsection, we focus on how to select the optimal relay $R$ and allocate the optimal transmit power $P_s$, $P_r$ and $P_z$ to maximize the instantaneous secrecy rate stated as (15) assuming the avaibility of global instantaneous CSI. The assumption is reasonable in some cases [34]-[36]. For example, we consider the scenario in which a legal user in the wireless network is captured by Trojan and slaved as an eavesdropper to tap the signal transmission. In this case, a broadcasting network consisting of multiple legal receivers among which some are slaved by Trojan and become eavesdroppers, turns into a wiretap system, where the global instantaneous CSI of the eavesdroppers may be available. For another example, in a scenario where a single active eavesdropper is registered in the wireless network as a subscribed user and exchanges signaling messages with the source and intermediate nodes, the global instantaneous CSI of the eavesdropper may be known.\\
\indent Under the FCSI assumption, the joint relay selection and power allocation problem can be formulated as
\begin{eqnarray}\label{(16)}
(R^*,P_s^*,P_r^*,P_z^*)&\!=&\mathrm{arg} \, \underset{i,P_s,P_r,P_z}{\mathrm{max}} \quad \frac {1+\gamma_d(i,P_s,P_i)} {1+\gamma_e(i,P_s,P_i,P_z)}\nonumber\\
&& {}
\begin{array}{lll}
\mathrm{s.t.} \, & P_s+P_r+P_z=P\\
{}& P_s \geq 0,P_r \geq 0,P_z \geq 0\\
{}& \log_2(1+\frac {|h_{si}|^2 P_s}{\delta_n^2}) \geq R_d\\
\end{array}
\end{eqnarray}
where $P$ is the total power budget for transmitting one source symbol, which is also the sum of the transmit power at the source, relay and jammers. Problem (16) can be performed in two steps. At the first step, each intermediate node $i$ in the decoding set $\cal{D}$ is treated as a candidate of relay. For each candidate $i \in \cal{D}$, we solve the following power allocation problem to maximize the secrecy rate of the corresponding JRJS scheme,
\begin{eqnarray}\label{(17)}
(P_{si}^*,P_{ri}^*,P_{zi}^*)&\!=&\mathrm{arg} \, \underset{P_s,P_r,P_z}{\mathrm{max}} \quad \frac {1+\gamma_d(i,P_s,P_r)} {1+\gamma_e(i,P_s,P_r,P_z)}\nonumber\\
&& {}
\begin{array}{lll}
\mathrm{s.t.} \, & P_s+P_r+P_z=P\\
{}& P_s \geq 0,P_r \geq 0,P_z \geq 0\\
{}&\log_2(1+\frac {|h_{si}|^2 P_s}{\delta_n^2}) \geq R_d .\\
\end{array}
\end{eqnarray}
At the second step, the candidate which leads to the maximum of the secrecy rate results with the optimal power allocation solution above is selected as the relay $R$. That is,
\begin{equation}\label{(18)}
R^*=\mathrm{arg} \, \underset{i\in \cal{D}}{\mathrm{max}} \quad \frac {1+\gamma_d(i,P_{si}^*,P_{ri}^*)} {1+\gamma_e(i,P_{si}^*,P_{ri}^*,P_{zi}^*)}.
\end{equation}
Then, the optimal power allocation solution of the proposed JRJS scheme is the solution of problem (17) when the candidate is $R^*$.\\
\indent It is noted that problem (18) can be easily solved by simple comparison. Therefore, we focus on how to solve problem (17). Without loss of generality, we denote the candidate of relay as $R$. Let us define $\lambda_e=\bm{h}_e^{\textrm{H}}\bm{z}\bm{z}^{\textrm{H}}\bm{h}_e$. Then, due to $P_z= P-P_s-P_r$,  problem (17) can be rewritten as (19).
\newcounter{TempEqCnt1}
\setcounter{TempEqCnt1}{19}
\setcounter{equation}{18}
\begin{figure*}
\begin{eqnarray}\label{(19)}
 (P_{sR}^*,P_{rR}^*)&\!=\!&\mathrm{arg} \ \underset{P_s,P_r}{\mathrm{max}} f(P_s,P_r)=\frac {  1+\mathrm{min}(\frac{|h_{sr}|^2 P_s}{\delta_n^2},\frac{|h_{rd}|^2 P_r}{\delta_n^2})  } {  1+\frac{|h_{se}|^2 P_s}{\delta_n^2}+\frac{|h_{re}|^2 P_r}{\delta_n^2+(P-P_s-P_r)\lambda_e}  }\nonumber\\
&& {}
\begin{array}{lll}
\mathrm{s.t.} \, &P_s+P_r \leq P\\
{}&P_s \geq 0,P_r \geq 0 \\
{}&\log_2(1+\frac {|h_{sr}|^2 P_s}{\delta_n^2}) \geq R_d .\\
\end{array}
\end{eqnarray}
\end{figure*}
\setcounter{equation}{19}

\indent Due to
\begin{equation}\label{(20)}
\mathrm{min}(\! \frac{|h_{sr}|^2 P_s}{\delta_n^2},\frac{|h_{rd}|^2 P_r}{\delta_n^2} \!)\!\!=\!\! \left \lbrace \!\!\!\!  \begin{array}{ll} \frac{|h_{sr}|^2 P_s}{\delta_n^2} & if \!\!\!\! \quad |h_{sr}|^2 P_s \leq |h_{rd}|^2 P_r \\ \frac{|h_{rd}|^2 P_r}{\delta_n^2} & if \!\!\!\! \quad |h_{sr}|^2 P_s \geq |h_{rd}|^2 P_r\end{array} \!\!\!,\\ \right.
\end{equation}
(19) can be broken down into two sub-problems, which are given by (21) and (22).
\newcounter{TempEqCnt2}
\setcounter{TempEqCnt2}{21}
\setcounter{equation}{20}
\begin{figure*}
\begin{eqnarray}\label{(21)}
(P_{sR}^*,P_{rR}^*)&\!=\!&\mathrm{arg} \ \underset{P_s,P_r} {\mathrm{max}} f_1(P_s,P_r)= \frac {  1+\frac{|h_{sr}|^2 P_s}{\delta_n^2}  } {  1+\frac{|h_{se}|^2 P_s}{\delta_n^2}+\frac{|h_{re}|^2 P_r}{\delta_n^2+(P-P_s-P_r)\lambda_e}  }\nonumber\\
&& {}
\begin{array}{lll}
\mathrm{s.t.} \,  & P_s+P_r \leq P\\
{}&P_s \geq 0,P_r \geq 0 \\
{}&\log_2(1+\frac {|h_{sr}|^2 P_s}{\delta_n^2}) \geq R_d \\
{}&|h_{sr}|^2 P_s \leq |h_{rd}|^2 P_r \\
\end{array}
\end{eqnarray}
\end{figure*}
\setcounter{equation}{21}

\newcounter{TempEqCnt3}
\setcounter{TempEqCnt3}{22}
\setcounter{equation}{21}
\begin{figure*}
\begin{eqnarray}\label{(22)}
(P_{sR}^*,P_{rR}^*)&\!=\!&\mathrm{arg} \ \underset{P_s,P_r} {\mathrm{max}} f_2(P_s,P_r)= \frac {1+\frac{|h_{rd}|^2 P_r}{\delta_n^2}  } {  1+\frac{|h_{se}|^2 P_s}{\delta_n^2}+\frac{|h_{re}|^2 P_r}{\delta_n^2+(P-P_s-P_r)\lambda_e}  }\nonumber\\
&& {}
\begin{array}{lll}
\mathrm{s.t.} \,  &P_s+P_r \leq P\\
{}&P_s \geq 0,P_r \geq 0 \\
{}&\log_2(1+\frac {|h_{sr}|^2 P_s}{\delta_n^2}) \geq R_d \\
{}&|h_{sr}|^2 P_s \geq |h_{rd}|^2 P_r .\\
\end{array}
\end{eqnarray}
\end{figure*}
\setcounter{equation}{22}

\indent As $f_1(P_s,P_r)$ is a decreasing function of $P_r$, $P_r$ should be minimized in order to maximize $f_1(P_s,P_r)$. Due to the condition $|h_{sr}|^2 P_s \leq |h_{rd}|^2 P_r$ in problem (21), we can easily obtain $P_r \geq \frac{|h_{sr}|^2}{|h_{rd}|^2 }P_s$. Thus, we set $P_r=\frac{|h_{sr}|^2}{|h_{rd}|^2 }P_s$ to obtain a sub-optimal solution of problem (21). Combining $P_r=\frac{|h_{sr}|^2}{|h_{rd}|^2 }P_s$ and $P_s+P_r \leq P$, we have $P_s \leq \frac{|h_{rd}|^2}{|h_{sr}|^2+|h_{rd}|^2}P$. Additionally, due to $\log_2(1+\frac {|h_{sr}|^2 P_s}{\delta_n^2}) \geq R_d$, we can obtain $P_s \geq \frac{2^{R_d}-1}{|h_{sr}|^2}\delta_n^2$. Then, problem (21) becomes (23) where
\newcounter{TempEqCnt4}
\setcounter{TempEqCnt4}{23}
\setcounter{equation}{22}
\begin{figure*}
\begin{eqnarray}\label{(23)}
P_{sR}^{\textrm{opt}}&\!=\!&\mathrm{arg} \ \underset{P_s} {\mathrm{max}} g_1(P_s)= \frac {1+\frac{|h_{sr}|^2 P_s}{\delta_n^2}  } {  1+\frac{|h_{se}|^2 P_s}{\delta_n^2}+\frac{|h_{re}|^2 \frac{|h_{sr}|^2}{|h_{rd}|^2 }P_s}{\delta_n^2+(P-P_s-\frac{|h_{sr}|^2}{|h_{rd}|^2 }P_s)\lambda_e}  }\nonumber\\
&& {}
\begin{array}{lll}
\mathrm{s.t.} \,  & P_{b1} \leq P_s \leq P_{b2}\\
\end{array}
\end{eqnarray}
\end{figure*}
\setcounter{equation}{23}
\begin{equation}\label{(24)}
P_{b1}=\frac{2^{R_d}-1}{|h_{sr}|^2}\delta_n^2
\end{equation}
and
\begin{equation}\label{(25)}
P_{b2}=\frac{|h_{rd}|^2}{|h_{sr}|^2+|h_{rd}|^2}P .
\end{equation}
Note that we should use
\begin{equation}\label{(26)}
R_d \leq \log_2(\frac{|h_{sr}|^2|h_{rd}|^2}{|h_{sr}|^2+|h_{rd}|^2}\frac{P}{\delta_n^2}+1)
\end{equation}
for signal transmission since $P_{b1} \leq P_{b2}$. The corresponding sub-optimal transmit power at $R$  and all the jammers is $P_{rR}^{\textrm{opt}}=\frac{|h_{sr}|^2}{|h_{rd}|^2 }P_{sR}^{\textrm{opt}}$ and $P_{zR}^{\textrm{opt}}=P-P_{sR}^{\textrm{opt}}-P_{rR}^{\textrm{opt}}$.\\
\indent For simplicity, define $\alpha_1=\frac{|h_{sr}|^2}{|h_{rd}|^2}+1$, $\beta_1=\frac{|h_{sr}|^2|h_{re}|^2}{|h_{rd}|^2}$, $\mu_1=P\lambda_e+\delta_n^2$, $\theta_1=\mu_1\delta_n^2$, $\phi_1=\mu_1|h_{sr}|^2-\alpha_1\delta_n^2\lambda_e$, $\tau_1=\alpha_1\lambda_e|h_{sr}|^2$, $\phi_2=\mu_1|h_{se}|^2-\alpha_1\delta_n^2\lambda_e+\beta_1\delta_n^2$ and $\tau_2=\alpha_1\lambda_e|h_{se}|^2$. Then, $g_1(P_s)$ can be rewritten as
\begin{equation}\label{(27)}
g_1(P_s)=\frac{\theta_1+\phi_1P_s-\tau_1P_s^2}{\theta_1+\phi_2P_s-\tau_2P_s^2}.
\end{equation}
\indent By taking the derivative of $g_1(P_s)$  with respect to $P_s$, we obtain
\begin{equation}\label{(28)}
\frac {dg_1(P_s)} {dP_s}=\frac {(\tau_2\phi_1-\tau_1\phi_2)P_s^2+2(\tau_2-\tau_1)\theta_1P_s+(\phi_1-\phi_2)\theta_1}{(\theta_1+\phi_2P_s-\tau_2P_s^2)^2}.
\end{equation}
As $|h_{se}|^2$ is not equal to $|h_{sr}|^2$ generally, it always satisfies that $\tau_2-\tau_1 \neq 0$. Thus, when $\tau_2\phi_1-\tau_1\phi_2=0$, the solution of $\frac {dg_1(P_s)} {dP_s}=0$ is easily obtained as
\begin{equation}\label{(29)}
P_{sR}^{\textrm{p-opt1}}=\frac{\phi_2-\phi_1}{2(\tau_2-\tau_1)}.
\end{equation}
Then, the solution of problem (23) is
\begin{equation}\label{(30)}
P_{sR}^{\textrm{opt1}}=\mathrm{arg} \, \underset{\substack{P_s \in \left\{P_{sR}^{\textrm{p-opt1}},P_{b1},P_{b2}\right\}\\   P_s \in \left[ P_{b1},P_{b2} \right]}} {\mathrm{max}} g_1(P_s)
\end{equation}
where $ \left\{x_1, x_2,...,x_n\right\}$ denotes a set consisting of elements $x_1, x_2,...,x_n$.\\
\indent Now we discuss the solution of problem (23) when $\tau_2\phi_1-\tau_1\phi_2 \neq 0$. Define $q_1(P_r)=(\tau_2\phi_1-\tau_1\phi_2)P_s^2+2(\tau_2-\tau_1)\theta_1P_s+(\phi_1-\phi_2)\theta_1$. The discriminant of $q_1(P_r)$ is obtained as
\begin{equation}\label{(31)}
\Delta_1=(2(\tau_2-\tau_1)\theta_1)^2-4(\tau_2\phi_1-\tau_1\phi_2)(\phi_1-\phi_2)\theta_1.
\end{equation}
When $\Delta_1>0$, the solutions of $\frac {dg_1(P_s)} {dP_s}=0$ are obtained as
\begin{equation}\label{(32)}
P_{sR}^{\textrm{p-opt2}}=\frac{-2(\tau_2-\tau_1)\theta_1+\sqrt{\Delta_1}}{2(\tau_2\phi_1-\tau_1\phi_2)}
\end{equation}
and
\begin{equation}\label{(33)}
P_{sR}^{\textrm{p-opt3}}=\frac{-2(\tau_2-\tau_1)\theta_1-\sqrt{\Delta_1}}{2(\tau_2\phi_1-\tau_1\phi_2)}.
\end{equation}
Then, the solution of problem (23) is
\begin{equation}\label{(34)}
P_{sR}^{\textrm{opt2}}=\mathrm{arg} \, \underset{\substack{P_s \in \left\{P_{sR}^{\textrm{p-opt2}},P_{sR}^{\textrm{p-opt3}},P_{b1},P_{b2}\right\}\\ P_s \in \left[ P_{b1},P_{b2} \right]}} {\mathrm{max}} g_1(P_s) .
\end{equation}
When $\Delta_1=0$, the solution of $\frac {dg_1(P_s)} {dP_s}=0$ is
\begin{equation}\label{(35)}
P_{sR}^{\textrm{p-opt4}}=\frac{(\tau_1-\tau_2)\theta_1}{\tau_2\phi_1-\tau_1\phi_2}.
\end{equation}
Then, the solution of problem (23) is
\begin{equation}\label{(36)}
P_{sR}^{\textrm{opt3}}=\mathrm{arg} \, \underset{\substack{P_s \in \left\{P_{sR}^{\textrm{p-opt4}},P_{b1},P_{b2}\right\}\\ P_s \in \left[ P_{b1},P_{b2} \right]}} {\mathrm{max}} g_1(P_s).
\end{equation}
When $\Delta_1<0$, $\frac {dg_1(P_s)} {dP_s}=0$ has no solutions. Then, the solution of problem (23) is
\begin{equation}\label{(37)}
P_{sR}^{\textrm{opt4}}=\mathrm{arg} \, \underset{P_s \in \left\{P_{b1},P_{b2}\right\}} {\mathrm{max}} g_1(P_s).
\end{equation}
\indent Next, we will discuss the solution of sub-problem (22) in a similar way. As $f_2(P_s,P_r)$ is a decreasing function of $P_s$, $P_s$ should be minimized in order to maximize $f_2(P_s,P_r)$. Due to the condition $|h_{sr}|^2 P_s \geq |h_{rd}|^2 P_r$ in problem (22), we set $P_s=\frac{|h_{rd}|^2}{|h_{sr}|^2 }P_r$ to obtain a sub-optimal solution of problem (22). Combining $P_s=\frac{|h_{rd}|^2}{|h_{sr}|^2 }P_r$ and $P_s+P_r \leq P$, we have $P_r \leq \frac{|h_{sr}|^2}{|h_{sr}|^2+|h_{rd}|^2}P$. Additionally, due to $P_s=\frac{|h_{rd}|^2}{|h_{sr}|^2 }P_r$ and $\log_2(1+\frac {|h_{sr}|^2 P_s}{\delta_n^2}) \geq R_d$, we can obtain $P_r \geq \frac{2^{R_d}-1}{|h_{rd}|^2}\delta_n^2$. Then, problem (22) becomes (38) where
\newcounter{TempEqCnt5}
\setcounter{TempEqCnt5}{38}
\setcounter{equation}{37}
\begin{figure*}
\begin{eqnarray}\label{(38)}
P_{rR}^{\textrm{opt}}&\!=\!&\mathrm{arg} \, \underset{P_r} {\mathrm{max}} g_2(P_r)= \frac {1+\frac{|h_{rd}|^2 P_r}{\delta_n^2}  } {  1+\frac{|h_{se}|^2 \frac{|h_{rd}|^2}{|h_{sr}|^2 }P_r}{\delta_n^2}+\frac{|h_{re}|^2 P_r}{\delta_n^2+(P-P_r-\frac{|h_{rd}|^2}{|h_{sr}|^2 }P_r)\lambda_e}  }\nonumber\\
&& {}
\begin{array}{lll}
\mathrm{s.t.} \,  & P_{b3} \leq P_r \leq P_{b4}\\
\end{array}
\end{eqnarray}
\end{figure*}
\setcounter{equation}{38}
\begin{equation}\label{(39)}
P_{b3}=\frac{2^{R_d}-1}{|h_{rd}|^2}\delta_n^2
\end{equation}
and
\begin{equation}\label{(40)}
P_{b4}=\frac{|h_{sr}|^2}{|h_{sr}|^2+|h_{rd}|^2}P .
\end{equation}
We can have that (26) still holds due to $P_{b3} \leq P_{b4}$. Then, the corresponding sub-optimal transmit power at the source and all the jammers is $P_{sR}^{\textrm{opt}}=\frac{|h_{rd}|^2}{|h_{sr}|^2 }P_{rR}^{\textrm{opt}}$ and $P_{zR}^{\textrm{opt}}=P-P_{sR}^{\textrm{opt}}-P_{rR}^{\textrm{opt}}$.\\
\indent For simplicity, define $\alpha_2=\frac{|h_{rd}|^2}{|h_{sr}|^2}+1$, $\beta_2=\frac{|h_{rd}|^2|h_{se}|^2}{|h_{sr}|^2}$, $\phi_3=\mu_1|h_{rd}|^2-\alpha_2\delta_n^2\lambda_e$, $\tau_3=\alpha_2\lambda_e|h_{rd}|^2$, $\phi_4=\mu_1\beta_2-\alpha_2\delta_n^2\lambda_e+|h_{re}|^2\delta_n^2$ and $\tau_4=\alpha_2\beta_2\lambda_e$. Then, $g_2(P_r)$ can be rewritten as
\begin{equation}\label{(41)}
g_2(P_r)=\frac{\theta_1+\phi_3P_r-\tau_3P_r^2}{\theta_1+\phi_4P_r-\tau_4P_r^2} .
\end{equation}
\indent By taking the derivative of $g_2(P_r)$  with respect to $P_r$, we obtain
\begin{equation}\label{(42)}
\frac {dg_2(P_r)} {dP_r}=\frac {(\tau_4\phi_3-\tau_3\phi_4)P_r^2+2(\tau_4-\tau_3)\theta_1P_r+(\phi_3-\phi_4)\theta_1}{(\theta_1+\phi_4P_r-\tau_4P_r^2)^2} .
\end{equation}
\indent As $|h_{se}|^2$ is not equal to $|h_{sr}|^2$ generally, it always satisfies that $\tau_4-\tau_3 \neq 0$. Thus, when $\tau_4\phi_3-\tau_3\phi_4=0$, the solution of $\frac {dg_2(P_r)} {dP_r}=0$ is obtained as
\begin{equation}\label{(43)}
P_{rR}^{\textrm{p-opt1}}=\frac{\phi_4-\phi_3}{2(\tau_4-\tau_3)} .
\end{equation}
Then, the solution of problem (38) is
\begin{equation}\label{(44)}
P_{rR}^{\textrm{opt1}}=\mathrm{arg} \, \underset{\substack{P_r \in \left\{P_{rR}^{\textrm{p-opt1}},P_{b3},P_{b4}\right\}\\ P_r \in \left[ P_{b3},P_{b4} \right]}} {\mathrm{max}} g_2(P_r).
\end{equation}
\indent Now we discuss the solution of problem (38) when $\tau_4\phi_3-\tau_3\phi_4 \neq 0$. Define $q_2(P_r)=(\tau_4\phi_3-\tau_3\phi_4)P_r^2+2(\tau_4-\tau_3)\theta_1P_r+(\phi_3-\phi_4)\theta_1$. The discriminant of $q_2(P_r)$ is obtained as
\begin{equation}\label{(45)}
\Delta_2=(2(\tau_4-\tau_3)\theta_1)^2-4(\tau_4\phi_3-\tau_3\phi_4)(\phi_3-\phi_4)\theta_1.
\end{equation}
When $\Delta_2>0$, the solutions of $\frac {dg_2(P_r)} {dP_r}=0$ are obtained as
\begin{equation}\label{(46)}
P_{rR}^{\textrm{p-opt2}}=\frac{-2(\tau_4-\tau_3)\theta_1+\sqrt{\Delta_2}}{2(\tau_4\phi_3-\tau_3\phi_4)}
\end{equation}
and
\begin{equation}\label{(47)}
P_{rR}^{\textrm{p-opt3}}=\frac{-2(\tau_4-\tau_3)\theta_1-\sqrt{\Delta_2}}{2(\tau_4\phi_3-\tau_3\phi_4)}.
\end{equation}
Then, the solution of problem (38) is
\begin{equation}\label{(48)}
P_{rR}^{\textrm{opt2}}=\mathrm{arg} \, \underset{\substack{P_r \in \left\{P_{rR}^{\textrm{p-opt2}},P_{rR}^{\textrm{p-opt3}},P_{b3},P_{b4}\right\}\\ P_r \in \left[ P_{b3},P_{b4} \right]}} {\mathrm{max}} g_2(P_r) .
\end{equation}
When $\Delta_2=0$, the solution of $\frac {dg_2(P_r)} {dP_r}=0$ is
\begin{equation}\label{(49)}
P_{rR}^{\textrm{p-opt4}}=\frac{(\tau_3-\tau_4)\theta_1}{\tau_4\phi_3-\tau_3\phi_4} .
\end{equation}
Then, the solution of problem (38) is
\begin{equation}\label{(50)}
P_{rR}^{\textrm{opt3}}=\mathrm{arg} \, \underset{\substack{P_r \in \left\{P_{rR}^{\textrm{p-opt4}},P_{b3},P_{b4}\right\}\\ P_s \in \left[ P_{b3},P_{b4} \right]}} {\mathrm{max}} g_2(P_r)
\end{equation}
When $\Delta_2<0$, $\frac {dg_2(P_r)} {dP_r}=0$ has no solutions. Then, the solution of problem (38) is
\begin{equation}\label{(51)}
P_{rR}^{\textrm{opt4}}=\mathrm{arg} \, \underset{P_r \in \left\{P_{b3},P_{b4}\right\}} {\mathrm{max}} g_2(P_r) .
\end{equation}
It is noted that problem (30), (34), (36), (37), (44), (48), (50) and (51) can be easily solved by simple comparison.\\
\indent In summary, the sub-optimal solution of problem (19), $(P_{sR}^*,P_{rR}^*)$, is stated in TABLE  $\textrm{\uppercase\expandafter{\romannumeral1}}$, and correspondingly, $P_{zR}^*=P-P_{sR}^*-P_{rR}^*$.
\begin{table*}
\centering
\caption{sub-optimal solution of problem (19)}
\begin{tabular}{|c|c|c|c|c|c|}
\hline
\multicolumn{5}{|c|}{condition}&sub-optimal solution\\
\hline
$\tau_2\phi_1-\tau_1\phi_2$&$\Delta_1$&$\tau_4\phi_3-\tau_3\phi_4$&$\Delta_2$&others&$(P_{sR}^*,P_{rR}^*)$\\
\hline
0&$-$&0&$-$&$f(P_{sR}^{\textrm{opt1}},\frac {|h_{sr}|^2}{|h_{rd}|^2}P_{sR}^{\textrm{opt1}}) \geq f( \frac{|h_{rd}|^2}{|h_{sr}|^2} P_{rR}^{\textrm{opt1}},P_{rR}^{\textrm{opt1}})$&\multirow{4}{*}{$( P_{sR}^{\textrm{opt1}},\frac {|h_{sr}|^2}{|h_{rd}|^2}P_{sR}^{\textrm{opt1}} )$}\\
\cline{1-5}
0&$-$&$\neq 0$&$>0$&$f(P_{sR}^{\textrm{opt1}},\frac {|h_{sr}|^2}{|h_{rd}|^2}P_{sR}^{\textrm{opt1}}) \geq f( \frac{|h_{rd}|^2}{|h_{sr}|^2} P_{rR}^{\textrm{opt2}},P_{rR}^{\textrm{opt2}})$&\\
\cline{1-5}
0&$-$&$\neq 0$&$0$&$f(P_{sR}^{\textrm{opt1}},\frac {|h_{sr}|^2}{|h_{rd}|^2}P_{sR}^{\textrm{opt1}}) \geq f( \frac{|h_{rd}|^2}{|h_{sr}|^2} P_{rR}^{\textrm{opt3}},P_{rR}^{\textrm{opt3}})$&\\
\cline{1-5}
0&$-$&$\neq 0$&$<0$&$f(P_{sR}^{\textrm{opt1}},\frac {|h_{sr}|^2}{|h_{rd}|^2}P_{sR}^{\textrm{opt1}}) \geq f( \frac{|h_{rd}|^2}{|h_{sr}|^2} P_{rR}^{\textrm{opt4}},P_{rR}^{\textrm{opt4}})$&\\
\hline
$\neq 0$&$>0$&0&$-$&$f(P_{sR}^{\textrm{opt2}},\frac {|h_{sr}|^2}{|h_{rd}|^2}P_{sR}^{\textrm{opt2}}) \geq f( \frac{|h_{rd}|^2}{|h_{sr}|^2} P_{rR}^{\textrm{opt1}},P_{rR}^{\textrm{opt1}})$&\multirow{4}{*}{$( P_{sR}^{\textrm{opt2}},\frac {|h_{sr}|^2}{|h_{rd}|^2}P_{sR}^{\textrm{opt2}} )$}\\
\cline{1-5}
$\neq 0$&$>0$&$\neq 0$&$>0$&$f(P_{sR}^{\textrm{opt2}},\frac {|h_{sr}|^2}{|h_{rd}|^2}P_{sR}^{\textrm{opt2}}) \geq f( \frac{|h_{rd}|^2}{|h_{sr}|^2} P_{rR}^{\textrm{opt2}},P_{rR}^{\textrm{opt2}})$&\\
\cline{1-5}
$\neq 0$&$>0$&$\neq 0$&$0$&$f(P_{sR}^{\textrm{opt2}},\frac {|h_{sr}|^2}{|h_{rd}|^2}P_{sR}^{\textrm{opt2}}) \geq f( \frac{|h_{rd}|^2}{|h_{sr}|^2} P_{rR}^{\textrm{opt3}},P_{rR}^{\textrm{opt3}})$&\\
\cline{1-5}
$\neq 0$&$>0$&$\neq 0$&$<0$&$f(P_{sR}^{\textrm{opt2}},\frac {|h_{sr}|^2}{|h_{rd}|^2}P_{sR}^{\textrm{opt2}}) \geq f( \frac{|h_{rd}|^2}{|h_{sr}|^2} P_{rR}^{\textrm{opt4}},P_{rR}^{\textrm{opt4}})$&\\
\hline
$\neq 0$&$0$&0&$-$&$f(P_{sR}^{\textrm{opt3}},\frac {|h_{sr}|^2}{|h_{rd}|^2}P_{sR}^{\textrm{opt3}}) \geq f( \frac{|h_{rd}|^2}{|h_{sr}|^2} P_{rR}^{\textrm{opt1}},P_{rR}^{\textrm{opt1}})$&\multirow{4}{*}{$( P_{sR}^{\textrm{opt3}},\frac {|h_{sr}|^2}{|h_{rd}|^2}P_{sR}^{\textrm{opt3}} )$}\\
\cline{1-5}
$\neq 0$&$0$&$\neq 0$&$>0$&$f(P_{sR}^{\textrm{opt3}},\frac {|h_{sr}|^2}{|h_{rd}|^2}P_{sR}^{\textrm{opt3}}) \geq f( \frac{|h_{rd}|^2}{|h_{sr}|^2} P_{rR}^{\textrm{opt2}},P_{rR}^{\textrm{opt2}})$&\\
\cline{1-5}
$\neq 0$&$0$&$\neq 0$&$0$&$f(P_{sR}^{\textrm{opt3}},\frac {|h_{sr}|^2}{|h_{rd}|^2}P_{sR}^{\textrm{opt3}}) \geq f( \frac{|h_{rd}|^2}{|h_{sr}|^2} P_{rR}^{\textrm{opt3}},P_{rR}^{\textrm{opt3}})$&\\
\cline{1-5}
$\neq 0$&$0$&$\neq 0$&$<0$&$f(P_{sR}^{\textrm{opt3}},\frac {|h_{sr}|^2}{|h_{rd}|^2}P_{sR}^{\textrm{opt3}}) \geq f( \frac{|h_{rd}|^2}{|h_{sr}|^2} P_{rR}^{\textrm{opt4}},P_{rR}^{\textrm{opt4}})$&\\
\hline
$\neq 0$&$<0$&0&$-$&$f(P_{sR}^{\textrm{opt4}},\frac {|h_{sr}|^2}{|h_{rd}|^2}P_{sR}^{\textrm{opt4}}) \geq f( \frac{|h_{rd}|^2}{|h_{sr}|^2} P_{rR}^{\textrm{opt1}},P_{rR}^{\textrm{opt1}})$&\multirow{4}{*}{$( P_{sR}^{\textrm{opt4}},\frac {|h_{sr}|^2}{|h_{rd}|^2}P_{sR}^{\textrm{opt4}} )$}\\
\cline{1-5}
$\neq 0$&$<0$&$\neq 0$&$>0$&$f(P_{sR}^{\textrm{opt4}},\frac {|h_{sr}|^2}{|h_{rd}|^2}P_{sR}^{\textrm{opt4}}) \geq f( \frac{|h_{rd}|^2}{|h_{sr}|^2} P_{rR}^{\textrm{opt2}},P_{rR}^{\textrm{opt2}})$&\\
\cline{1-5}
$\neq 0$&$<0$&$\neq 0$&$0$&$f(P_{sR}^{\textrm{opt4}},\frac {|h_{sr}|^2}{|h_{rd}|^2}P_{sR}^{\textrm{opt4}}) \geq f( \frac{|h_{rd}|^2}{|h_{sr}|^2} P_{rR}^{\textrm{opt3}},P_{rR}^{\textrm{opt3}})$&\\
\cline{1-5}
$\neq 0$&$<0$&$\neq 0$&$<0$&$f(P_{sR}^{\textrm{opt4}},\frac {|h_{sr}|^2}{|h_{rd}|^2}P_{sR}^{\textrm{opt4}}) \geq f( \frac{|h_{rd}|^2}{|h_{sr}|^2} P_{rR}^{\textrm{opt4}},P_{rR}^{\textrm{opt4}})$&\\
\hline
0&$-$&0&$-$&$f(P_{sR}^{\textrm{opt1}},\frac {|h_{sr}|^2}{|h_{rd}|^2}P_{sR}^{\textrm{opt1}}) \leq f( \frac{|h_{rd}|^2}{|h_{sr}|^2} P_{rR}^{\textrm{opt1}},P_{rR}^{\textrm{opt1}})$&\multirow{4}{*}{$( \frac{|h_{rd}|^2}{|h_{sr}|^2} P_{rR}^{\textrm{opt1}},P_{rR}^{\textrm{opt1}} )$}\\
\cline{1-5}
$\neq 0$&$>0$&0&$-$&$f(P_{sR}^{\textrm{opt2}},\frac {|h_{sr}|^2}{|h_{rd}|^2}P_{sR}^{\textrm{opt2}}) \leq f( \frac{|h_{rd}|^2}{|h_{sr}|^2} P_{rR}^{\textrm{opt1}},P_{rR}^{\textrm{opt1}})$&\\
\cline{1-5}
$\neq 0$&$0$&0&$-$&$f(P_{sR}^{\textrm{opt3}},\frac {|h_{sr}|^2}{|h_{rd}|^2}P_{sR}^{\textrm{opt3}}) \leq f( \frac{|h_{rd}|^2}{|h_{sr}|^2} P_{rR}^{\textrm{opt1}},P_{rR}^{\textrm{opt1}})$&\\
\cline{1-5}
$\neq 0$&$<0$&0&$-$&$f(P_{sR}^{\textrm{opt4}},\frac {|h_{sr}|^2}{|h_{rd}|^2}P_{sR}^{\textrm{opt4}}) \leq f( \frac{|h_{rd}|^2}{|h_{sr}|^2} P_{rR}^{\textrm{opt1}},P_{rR}^{\textrm{opt1}})$&\\
\hline
0&$-$&$\neq 0$&$>0$&$f(P_{sR}^{\textrm{opt1}},\frac {|h_{sr}|^2}{|h_{rd}|^2}P_{sR}^{\textrm{opt1}}) \leq f( \frac{|h_{rd}|^2}{|h_{sr}|^2} P_{rR}^{\textrm{opt2}},P_{rR}^{\textrm{opt2}})$&\multirow{4}{*}{$( \frac{|h_{rd}|^2}{|h_{sr}|^2} P_{rR}^{\textrm{opt2}},P_{rR}^{\textrm{opt2}} )$}\\
\cline{1-5}
$\neq 0$&$>0$&$\neq 0$&$>0$&$f(P_{sR}^{\textrm{opt2}},\frac {|h_{sr}|^2}{|h_{rd}|^2}P_{sR}^{\textrm{opt2}}) \leq f( \frac{|h_{rd}|^2}{|h_{sr}|^2} P_{rR}^{\textrm{opt2}},P_{rR}^{\textrm{opt2}})$&\\
\cline{1-5}
$\neq 0$&$0$&$\neq 0$&$>0$&$f(P_{sR}^{\textrm{opt3}},\frac {|h_{sr}|^2}{|h_{rd}|^2}P_{sR}^{\textrm{opt3}}) \leq f( \frac{|h_{rd}|^2}{|h_{sr}|^2} P_{rR}^{\textrm{opt2}},P_{rR}^{\textrm{opt2}})$&\\
\cline{1-5}
$\neq 0$&$<0$&$\neq 0$&$>0$&$f(P_{sR}^{\textrm{opt4}},\frac {|h_{sr}|^2}{|h_{rd}|^2}P_{sR}^{\textrm{opt4}}) \leq f( \frac{|h_{rd}|^2}{|h_{sr}|^2} P_{rR}^{\textrm{opt2}},P_{rR}^{\textrm{opt2}})$&\\
\hline
0&$-$&$\neq 0$&$0$&$f(P_{sR}^{\textrm{opt1}},\frac {|h_{sr}|^2}{|h_{rd}|^2}P_{sR}^{\textrm{opt1}}) \leq f( \frac{|h_{rd}|^2}{|h_{sr}|^2} P_{rR}^{\textrm{opt3}},P_{rR}^{\textrm{opt3}})$&\multirow{4}{*}{$( \frac{|h_{rd}|^2}{|h_{sr}|^2} P_{rR}^{\textrm{opt3}},P_{rR}^{\textrm{opt3}})$}\\
\cline{1-5}
$\neq 0$&$>0$&$\neq 0$&$0$&$f(P_{sR}^{\textrm{opt2}},\frac {|h_{sr}|^2}{|h_{rd}|^2}P_{sR}^{\textrm{opt2}}) \leq f( \frac{|h_{rd}|^2}{|h_{sr}|^2} P_{rR}^{\textrm{opt3}},P_{rR}^{\textrm{opt3}})$&\\
\cline{1-5}
$\neq 0$&$0$&$\neq 0$&$0$&$f(P_{sR}^{\textrm{opt3}},\frac {|h_{sr}|^2}{|h_{rd}|^2}P_{sR}^{\textrm{opt3}}) \leq f( \frac{|h_{rd}|^2}{|h_{sr}|^2} P_{rR}^{\textrm{opt3}},P_{rR}^{\textrm{opt3}})$&\\
\cline{1-5}
$\neq 0$&$<0$&$\neq 0$&$0$&$f(P_{sR}^{\textrm{opt4}},\frac {|h_{sr}|^2}{|h_{rd}|^2}P_{sR}^{\textrm{opt4}}) \leq f( \frac{|h_{rd}|^2}{|h_{sr}|^2} P_{rR}^{\textrm{opt3}},P_{rR}^{\textrm{opt3}})$&\\
\hline
0&$-$&$\neq 0$&$<0$&$f(P_{sR}^{\textrm{opt1}},\frac {|h_{sr}|^2}{|h_{rd}|^2}P_{sR}^{\textrm{opt1}}) \leq f( \frac{|h_{rd}|^2}{|h_{sr}|^2} P_{rR}^{\textrm{opt4}},P_{rR}^{\textrm{opt4}})$&\multirow{4}{*}{$( \frac{|h_{rd}|^2}{|h_{sr}|^2} P_{rR}^{\textrm{opt4}},P_{rR}^{\textrm{opt4}} )$}\\
\cline{1-5}
$\neq 0$&$>0$&$\neq 0$&$<0$&$f(P_{sR}^{\textrm{opt2}},\frac {|h_{sr}|^2}{|h_{rd}|^2}P_{sR}^{\textrm{opt2}}) \leq f( \frac{|h_{rd}|^2}{|h_{sr}|^2} P_{rR}^{\textrm{opt4}},P_{rR}^{\textrm{opt4}})$&\\
\cline{1-5}
$\neq 0$&$0$&$\neq 0$&$<0$&$f(P_{sR}^{\textrm{opt3}},\frac {|h_{sr}|^2}{|h_{rd}|^2}P_{sR}^{\textrm{opt3}}) \leq f( \frac{|h_{rd}|^2}{|h_{sr}|^2} P_{rR}^{\textrm{opt4}},P_{rR}^{\textrm{opt4}})$&\\
\cline{1-5}
$\neq 0$&$<0$&$\neq 0$&$<0$&$f(P_{sR}^{\textrm{opt4}},\frac {|h_{sr}|^2}{|h_{rd}|^2}P_{sR}^{\textrm{opt4}}) \leq f( \frac{|h_{rd}|^2}{|h_{sr}|^2} P_{rR}^{\textrm{opt4}},P_{rR}^{\textrm{opt4}})$&\\
\hline
\end{tabular}
\end{table*}

\subsection{PCSI based Power Allocation (PCSI-PA) Scheme}
Considering that it is generally difficult to obtain the instantaneous CSI of the wiretap channel in many practical cases, we propose a PCSI based power allocation (PCSI-PA) scheme in this subsection. We assume that the fading coefficient $h_{se}$ and $h_{ie}$ are complex Gaussian randoms with zero mean value and variance $\epsilon_1$ and $\epsilon_2$, respectively, that is, $\mathrm E \left[h_{se}\right]=\mathrm E \left[h_{ie}\right]=0$, $\mathrm E \left[|h_{se}|^2 \right] =\delta_{se}^2 =\epsilon_1$ and $\mathrm E \left[|h_{ie}|^2 \right] = \delta_{ie}^2=\epsilon_2$.\\
\indent As the CSI of the wiretap channel is unavailable, it is infeasible to maximize the instantaneous secrecy rate stated in (15). Thus, we aim to maximize the average secrecy rate of the source-destination transmission $\mathrm E \left[C_s \right]=\mathrm E \left[C_d-C_e \right]$. Due to
\begin{align}\label{(52)}
\mathrm E \left[C_d-C_e \right]&=\lim\limits_{L\rightarrow{\infty}} \frac {1}{L} \sum_{l=1}^L \left[ {C_{dl}(l)-C_{el}(l)} \right]\\ \nonumber
&=\lim\limits_{L\rightarrow{\infty}} \frac {1}{L} \sum_{l=1}^L \left[ {C_{dl}(l)-\rho(l)\mathrm E \left[C_e\right]} \right] \nonumber
\end{align}
where $C_{dl}(l)$ and $C_{el}(l)$ denote the instantaneous channel capacity of the main link and wiretap link at time $l$, respectively, and $\rho(l)$ is a weighting factor satisfying that $\rho(l)\geq 0, \sum\limits_{l=1}^L \rho(l)=L$. Thus, (52) can be optimized by maximizing $C_{dl}(l)-\rho(l)\mathrm E \left[C_e\right]$ for each $l$ by finding the optimal $\rho(l)$ and the optimal transmit power at the source and relay. Since it is difficult to obtain the optimal value of $\rho(l)$ for each $l$, we turn to use $C_{dl}(l)-\mathrm E \left[C_e\right]$ as an objective function instead by simply setting $\rho(l)=1$ for each $l$. As the optimization of $\rho(l)$ is not taken into account, we aim to obtain a lower bound of $\mathrm E \left[C_d-C_e \right]$ by using the objective function $C_{dl}(l)-\mathrm E \left[C_e\right]$. \\
\indent For simplicity, we denote $C_{dl}(l)$ uniformly as $C_d$ without loss of generality. As $\log_2(x)$ is a concave function of $x$, due to Jensen's inequality, we have
\begin{align}\label{(53)}
\mathrm E \left[C_e\right]&=\mathrm E \left[\log_2(1+\gamma_e(i,P_s,P_r,P_z))\right] \\ \nonumber
&\leq \log_2(1+\bar{\gamma_e}(i,P_s,P_r,P_z))
\end{align}
where $\bar{\gamma}_e(i,P_s,P_r,P_z)$ is expressed as
\begin{align}\label{(54)}
\bar{\gamma}_e(i,P_s,P_r,P_z)&=\mathrm E \left[\gamma_e(i,P_s,P_r,P_z)\right]\\ \nonumber
&=\frac {\mathrm E_{h_{se}} \left[|h_{se}|^2 \right] P_s}{\delta_n^2}+\mathrm E \left[ \frac {|h_{re}|^2 P_r} {\delta_n^2+P_z \bm{h}_e^{\textrm{H}}\bm{z}\bm{z}^{\textrm{H}}\bm{h}_e} \right]\\ \nonumber
&=\frac {\epsilon_1 P_s}{\delta_n^2}+\mathrm E \left[ \frac {|h_{re}|^2 P_r} {\delta_n^2+P_z \lambda_e} \right] \nonumber .
\end{align}\\
\indent Defining $e_\mathrm E=\frac{\mathrm E \left[ \frac {1} {\delta_n^2+P_z \lambda_e} \right] - \frac {1} {\delta_n^2+P_z \epsilon_2}}{\frac {1} {\delta_n^2+P_z \epsilon_2}}$, we can represent $e_\mathrm E$ as
\begin{align}\label{(55)}
e_\mathrm E&=\frac {\mathrm E \left[ \frac {1} {\delta_n^2+P_z \lambda_e} - \frac {1} {\delta_n^2+P_z \epsilon_2} \right]} {\frac {1} {\delta_n^2+P_z \epsilon_2}}\\ \nonumber
&=\mathrm E \left[ \frac {\epsilon_2- \lambda_e}{\frac{\delta_n^2}{P_z}+\lambda_e} \right]\nonumber .
\end{align}
Moreover, we can obtain the mean of $\lambda_e$ as
\begin{align}\label{(56)}
\mathrm E \left[ \lambda_e \right] &= \mathrm E \left[ \bm{h}_e^{\textrm{H}}\bm{z}\bm{z}^{\textrm{H}}\bm{h}_e \right]= \mathrm E \left[ \sum_{i=1}^{M-1} h_{j_ie} z_i \sum_{i=1}^{M-1} (h_{j_ie} z_i)^{\textrm{H}} \right]\\ \nonumber
&= \mathrm E \!\!\left[ \sum_{i=1}^{M-1} |h_{j_ie}|^2 |z_i|^2\!\!+\!\!\! \sum_{i_1=1}^{M-1} \!\! \sum_{i_2=1, i_2 \neq i_1}^{M-1} \!\!\!\!\!\!\!(h_{j_{i_1}e} z_{i_1}) (h_{j_{i_2}e} z_{i_2})^{\textrm{H}}  \right]  \\ \nonumber
&=\mathrm E \left[ \sum_{i=1}^{M-1} |h_{j_ie}|^2 |z_i|^2 \right] =\sum_{i=1}^{M-1} \mathrm E \left[ |h_{j_ie}|^2 \right] |z_i|^2  \\ \nonumber
&=\epsilon_2 \sum_{i=1}^{M-1} |z_i|^2=\epsilon_2   \nonumber .
\end{align}
It can be observed from (55) and (56) that, when the variance of $\lambda_e$ tends to 0, the random variable $\lambda_e$ converges to its mean $\epsilon_2$, leading to the fact that $\mathrm E \left[ \frac {1} {\delta_n^2+P_z \lambda_e} \right]$ approaches $\frac {1} {\delta_n^2+P_z \epsilon_2}$. Fig. 2 illustrates the values of $e_\mathrm E$ versus the variance of $\lambda_e$ in the cases of $\frac{P}{\delta_n^2}=0$ dB, $\frac{P}{\delta_n^2}=10$ dB and $\frac{P}{\delta_n^2}=20$ dB. In the simulation, $P_z$ is set to a random value in $\left[0, P \right]$. \\
\begin{figure}[htb!]
\centering
\includegraphics[scale=0.65]{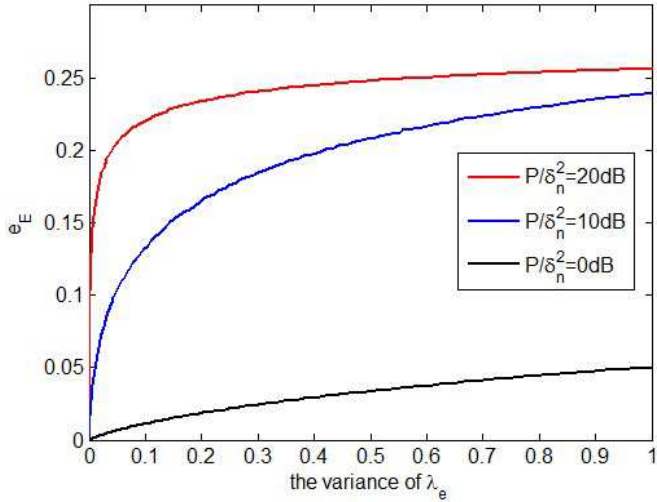}
\caption{An illustrstion of the values of $e_\mathrm E$ versus the variance of $\lambda_e$ in the cases of $\frac{P}{\delta_n^2}=0$ dB, $\frac{P}{\delta_n^2}=10$ dB and $\frac{P}{\delta_n^2}=20$ dB.}
\label{fig_2}
\end{figure}
\indent Thus, when the variance of $\lambda_e$ is small, $\mathrm E \left[ \frac {1} {\delta_n^2+P_z \lambda_e} \right]$ can be approximated by $\frac {1} {\delta_n^2+P_z \epsilon_2}$. Then, we have
\begin{align}\label{(57)}
\mathrm E \left[ \frac {|h_{re}|^2 P_r} {\delta_n^2+P_z \lambda_e} \right]&=\mathrm E \left[|h_{re}|^2 P_r  \right] \mathrm E \left[ \frac {1} {\delta_n^2+P_z \lambda_e} \right] \\ \nonumber
&\approx \mathrm E \left[|h_{re}|^2 P_r  \right] \frac {1} {\delta_n^2+P_z \epsilon_2}\\ \nonumber
&= \frac {\epsilon_2 P_r} {\delta_n^2+P_z \epsilon_2} \nonumber .
\end{align}
Combining (54) and (57), $\bar{\gamma}_e(i,P_s,P_r,P_z)$ can be approximately expressed as
\begin{equation}\label{(58)}
\bar{\gamma}_e(i,P_s,P_r,P_z) \approx \tilde{\gamma_e}=\frac {\epsilon_1 P_s}{\delta_n^2}+\frac {\epsilon_2 P_r} {\delta_n^2+P_z \epsilon_2}
\end{equation}
when the variance of $\lambda_e$ is small.\\
\indent From (53) and (58), we have
\begin{align}\label{(59)}
C_d-\mathrm E \left[C_e\right]&\geq C_d-\log_2(1+\tilde{\gamma_e}(i,P_s,P_r,P_z)) \\ \nonumber
&=\log_2 (\frac{1+\gamma_d (i,P_s,P_r)}{1+\tilde{\gamma_e}(i,P_s,P_r,P_z)})  .
\end{align}
Therefore, we aim to maximize the lower bound of $C_d-\mathrm E \left[C_e\right]$, that is, $\log_2 (\frac{1+\gamma_d (i,P_s,P_r)}{1+\tilde{\gamma_e}(i,P_s,P_r,P_z)})$, which is proved to be effective in our experiments. \\
\indent The PCSI-PA problem can be formulated as
\begin{eqnarray}\label{(60)}
(R^*,P_s^*,P_r^*,P_z^*)&\!=&\mathrm{arg} \, \underset{i,P_s,P_r,P_z}{\mathrm{max}} \quad \frac {1+\gamma_d(i,P_s,P_r)} {1+\tilde{\gamma_e}(i,P_s,P_r,P_z)}\nonumber\\
&& {}
\begin{array}{lll}
\mathrm{s.t.} \, & P_s+P_r+P_z=P\\
{}& P_s \geq 0,P_r \geq 0,P_z \geq 0\\
{}& \log_2(1+\frac {|h_{si}|^2 P_s}{\delta_n^2}) \geq R_d\\
\end{array}
\end{eqnarray}
which can be solved in a similar way as problem (16).\\
\indent For each candidate of relay $i \in \cal{D}$, the power allocation problem is formulated as
\begin{eqnarray}\label{(61)}
(P_{si}^*,P_{ri}^*,P_{zi}^*)&\!=&\mathrm{arg} \ \underset{P_s,P_r,P_z} {\mathrm{max}} \quad \frac {1+\gamma_d(i,P_s,P_r)} {1+\tilde{\gamma_e}(i,P_s,P_r,P_z)}\nonumber\\
&& {}
\begin{array}{lll}
\mathrm{s.t.} \, & P_s+P_r+P_z=P\\
{}& P_s \geq 0,P_r \geq 0,P_z \geq 0\\
{}& \log_2(1+\frac {|h_{si}|^2 P_s}{\delta_n^2}) \geq R_d\\
\end{array}
\end{eqnarray}
and the relay is determined as
\begin{equation}\label{(62)}
R^*=\mathrm{arg} \ \underset{i\in \cal{D}}{\mathrm{max}} \quad \frac {1+\gamma_d(i,P_{si}^*,P_{ri}^*)} {1+\tilde{\gamma_e}(i,P_{si}^*,P_{ri}^*,P_{zi}^*)}\\
\end{equation}
which can be also easily solved by simple comparison. \\
\indent Since $P_s+P_r+P_z=P$, problem (61) can be rewritten as
\begin{eqnarray}\label{(63)}
 [P_{sR}^*,P_{rR}^*]\!\!\!\!\!\!\!&=\!\!\!\!\!\!\!\!&\mathrm{arg} \ \underset{P_s,P_r} {\mathrm{max}} \widetilde f(P_s,P_r) \!\!=\!\! \frac {  1+\mathrm{min}(\frac{|h_{sr}|^2 P_s}{\delta_n^2},\frac{|h_{rd}|^2 P_r}{\delta_n^2})  } {  1+\frac{\epsilon_1 P_s}{\delta_n^2}+\frac{\epsilon_2 P_r}{\delta_n^2+(P-P_s-P_r)\epsilon_2}  }\nonumber\\
&& {}
\begin{array}{lll}
\mathrm{s.t.} \, &P_s+P_r \leq P\\
{}&P_s \geq 0,P_r \geq 0 \\
{}&\log_2(1+\frac {|h_{sr}|^2 P_s}{\delta_n^2}) \geq R_d .\\
\end{array}
\end{eqnarray}
We can obtain a sub-optimal solution of problem (63) in a similar way as that in Section $\textrm{\uppercase\expandafter{\romannumeral3}}$.A.\\
\indent Define $\beta_3=\frac{|h_{sr}|^2\epsilon_2}{|h_{rd}|^2}$, $\mu_2=P\epsilon_2+\delta_n^2$, $\theta_2=\mu_2\delta_n^2$, $\phi_5=\mu_2|h_{sr}|^2-\alpha_1\delta_n^2\epsilon_2$, $\tau_5=\alpha_1\epsilon_2|h_{sr}|^2$, $\phi_6=\mu_2\epsilon_1-\alpha_1\delta_n^2\epsilon_2+\beta_3\delta_n^2$, $\tau_6=\alpha_1\epsilon_1\epsilon_2$, $\Delta_3=4(\tau_6-\tau_5)^2\theta_2^2-4(\tau_6\phi_5-\tau_5\phi_6)(\phi_5-\phi_6)\theta_2$ and $g_3(P_s)=\frac{\theta_2+\phi_5P_s-\tau_5P_s^2}{\theta_2+\phi_6P_s-\tau_6P_s^2}$. Then, we have
\begin{equation}\label{(64)}
P_{sR}^{\textrm{p-opt5}}=\frac{\phi_6-\phi_5}{2(\tau_6-\tau_5)} ,
\end{equation}
\begin{equation}\label{(65)}
P_{sR}^{\textrm{opt5}}=\mathrm{arg} \, \underset{\substack{P_s \in \left\{P_s^{\textrm{p-opt5}},P_{b1},P_{b2}\right\}\\ P_s \in \left[ P_{b1},P_{b2} \right]}} {\mathrm{max}} g_3(P_s) ,
\end{equation}
\begin{equation}\label{(66)}
P_{sR}^{\textrm{p-opt6}}=\frac{-2(\tau_6-\tau_5)\theta_2+\sqrt{\Delta_3}}{2(\tau_6\phi_5-\tau_5\phi_6)} ,
\end{equation}
\begin{equation}\label{(67)}
P_{sR}^{\textrm{p-opt7}}=\frac{-2(\tau_6-\tau_5)\theta_2-\sqrt{\Delta_3}}{2(\tau_6\phi_5-\tau_5\phi_6)} ,
\end{equation}
\begin{equation}\label{(68)}
P_{sR}^{\textrm{opt6}}=\mathrm{arg} \, \underset{\substack{P_s \in \left\{P_{sR}^{\textrm{p-opt6}},P_{sR}^{\textrm{p-opt7}}, P_{b1},P_{b2}\right\}\\ P_s \in \left[ P_{b1},P_{b2} \right]}} {\mathrm{max}} g_3(P_s)     ,
\end{equation}
\begin{equation}\label{(69)}
P_{sR}^{\textrm{p-opt8}}=\frac{(\tau_5-\tau_6)\theta_2}{\tau_6\phi_5-\tau_5\phi_6}    ,
\end{equation}
\begin{equation}\label{(70)}
P_{sR}^{\textrm{opt7}}=\mathrm{arg} \, \underset{\substack{P_s \in \left\{P_{sR}^{\textrm{p-opt8}},P_{b1},P_{b2}\right\}\\ P_s \in \left[ P_{b1},P_{b2} \right]}} {\mathrm{max}} g_3(P_s) ,
\end{equation}
and
\begin{equation}\label{(71)}
P_{sR}^{\textrm{opt8}}=\mathrm{arg} \, \underset{P_s \in \left\{P_{b1},P_{b2}\right\}} {\mathrm{max}} g_3(P_s) .
\end{equation}
\indent Define $\beta_4=\frac{|h_{rd}|^2\epsilon_1}{|h_{sr}|^2}$, $\phi_7=\mu_2|h_{rd}|^2-\alpha_2\delta_n^2\epsilon_2$, $\tau_7=\alpha_2\epsilon_2|h_{rd}|^2$, $\phi_8=\beta_4\mu_2-\alpha_2\delta_n^2\epsilon_2+\epsilon_2\delta_n^2$, $\tau_8=\alpha_2\epsilon_2\beta_4$, $\Delta_4=4(\tau_8-\tau_7)^2\theta_2^2-4(\tau_8\phi_7-\tau_7\phi_8)(\phi_7-\phi_8)\theta_2$ and $g_4(P_r)=\frac{\theta_2+\phi_7P_r-\tau_7P_r^2}{\theta_2+\phi_8P_r-\tau_8P_r^2}$. Then, we have
\begin{equation}\label{(72)}
P_{rR}^{\textrm{p-opt5}}=\frac{\phi_8-\phi_7}{2(\tau_8-\tau_7)} ,
\end{equation}
\begin{equation}\label{(73)}
P_{rR}^{\textrm{opt5}}=\mathrm{arg} \, \underset{\substack{P_r \in \left\{P_r^{\textrm{p-opt5}},P_{b3},P_{b4}\right\}\\ P_r \in \left[ P_{b3},P_{b4} \right]}} {\mathrm{max}} g_4(P_r) ,
\end{equation}
\begin{equation}\label{(74)}
P_{rR}^{\textrm{p-opt6}}=\frac{-2(\tau_8-\tau_7)\theta_2+\sqrt{\Delta_4}}{2(\tau_8\phi_7-\tau_7\phi_8)} ,
\end{equation}
\begin{equation}\label{(75)}
P_{rR}^{\textrm{p-opt7}}=\frac{-2(\tau_8-\tau_7)\theta_2-\sqrt{\Delta_4}}{2(\tau_8\phi_7-\tau_7\phi_8)}  ,
\end{equation}
\begin{equation}\label{(76)}
P_{rR}^{\textrm{opt6}}=\mathrm{arg} \, \underset{\substack{P_r \in \left\{P_r^{\textrm{p-opt6}},P_r^{\textrm{p-opt7}},P_{b3},P_{b4}\right\}\\ P_r \in \left[ P_{b3},P_{b4} \right]}} {\mathrm{max}} g_4(P_r) ,
\end{equation}
\begin{equation}\label{(77)}
P_{rR}^{\textrm{p-opt8}}=\frac{(\tau_7-\tau_8)\theta_2}{\tau_8\phi_7-\tau_7\phi_8}  ,
\end{equation}
\begin{equation}\label{(78)}
P_{rR}^{\textrm{opt7}}=\mathrm{arg} \, \underset{\substack{P_r \in \left\{P_r^{\textrm{p-opt8}},P_{b3},P_{b4}\right\}\\ P_s \in \left[ P_{b3},P_{b4} \right]}} {\mathrm{max}} g_4(P_r)
\end{equation}
and
\begin{equation}\label{(79)}
P_{rR}^{\textrm{opt8}}=\mathrm{arg} \, \underset{P_r \in \left\{P_{b3},P_{b4}\right\}} {\mathrm{max}} g_4(P_r) .
\end{equation}
\indent We state the sub-optimal solution of problem (63) in TABLE $\textrm{\uppercase\expandafter{\romannumeral2}}$  and then the sub-optimal total transmit power at all the jammers $P_{zR}^*$ is $P-P_{sR}^*-P_{rR}^*$.
\begin{table*}
\centering
\caption{sub-optimal solution of problem (63)}
\begin{tabular}{|c|c|c|c|c|c|}
\hline
\multicolumn{5}{|c|}{condition}&sub-optimal solution\\
\hline
$\tau_6\phi_5-\tau_5\phi_6$&$\Delta_3$&$\tau_8\phi_7-\tau_7\phi_8$&$\Delta_4$&others&$[P_{sR}^*,P_{rR}^*]$\\
\hline
0&$-$&0&$-$&$\widetilde f(P_{sR}^{\textrm{opt5}},\frac {|h_{sr}|^2}{|h_{rd}|^2}P_{sR}^{\textrm{opt5}}) \geq \widetilde f( \frac{|h_{rd}|^2}{|h_{sr}|^2} P_{rR}^{\textrm{opt5}},P_{rR}^{\textrm{opt5}})$&\multirow{4}{*}{$\left[ P_{sR}^{\textrm{opt5}},\frac {|h_{sr}|^2}{|h_{rd}|^2}P_{sR}^{\textrm{opt5}} \right]$}\\
\cline{1-5}
0&$-$&$\neq 0$&$>0$&$\widetilde f(P_{sR}^{\textrm{opt5}},\frac {|h_{sr}|^2}{|h_{rd}|^2}P_{sR}^{\textrm{opt5}}) \geq \widetilde f( \frac{|h_{rd}|^2}{|h_{sr}|^2} P_{rR}^{\textrm{opt6}},P_{rR}^{\textrm{opt6}})$&\\
\cline{1-5}
0&$-$&$\neq 0$&$0$&$\widetilde f(P_{sR}^{\textrm{opt5}},\frac {|h_{sr}|^2}{|h_{rd}|^2}P_{sR}^{\textrm{opt5}}) \geq \widetilde f( \frac{|h_{rd}|^2}{|h_{sr}|^2} P_{rR}^{\textrm{opt7}},P_{rR}^{\textrm{opt7}})$&\\
\cline{1-5}
0&$-$&$\neq 0$&$<0$&$\widetilde f(P_{sR}^{\textrm{opt5}},\frac {|h_{sr}|^2}{|h_{rd}|^2}P_{sR}^{\textrm{opt5}}) \geq \widetilde f( \frac{|h_{rd}|^2}{|h_{sr}|^2} P_{rR}^{\textrm{opt8}},P_{rR}^{\textrm{opt8}})$&\\
\hline
$\neq 0$&$>0$&0&$-$&$\widetilde f(P_{sR}^{\textrm{opt6}},\frac {|h_{sr}|^2}{|h_{rd}|^2}P_{sR}^{\textrm{opt6}}) \geq \widetilde f( \frac{|h_{rd}|^2}{|h_{sr}|^2} P_{rR}^{\textrm{opt5}},P_{rR}^{\textrm{opt5}})$&\multirow{4}{*}{$\left[ P_{sR}^{\textrm{opt6}},\frac {|h_{sr}|^2}{|h_{rd}|^2}P_{sR}^{\textrm{opt6}} \right]$}\\
\cline{1-5}
$\neq 0$&$>0$&$\neq 0$&$>0$&$\widetilde f(P_{sR}^{\textrm{opt6}},\frac {|h_{sr}|^2}{|h_{rd}|^2}P_{sR}^{\textrm{opt6}}) \geq \widetilde f( \frac{|h_{rd}|^2}{|h_{sr}|^2} P_{rR}^{\textrm{opt6}},P_{rR}^{\textrm{opt6}})$&\\
\cline{1-5}
$\neq 0$&$>0$&$\neq 0$&$0$&$\widetilde f(P_{sR}^{\textrm{opt6}},\frac {|h_{sr}|^2}{|h_{rd}|^2}P_{sR}^{\textrm{opt6}}) \geq \widetilde f( \frac{|h_{rd}|^2}{|h_{sr}|^2} P_{rR}^{\textrm{opt7}},P_{rR}^{\textrm{opt7}})$&\\
\cline{1-5}
$\neq 0$&$>0$&$\neq 0$&$<0$&$\widetilde f(P_{sR}^{\textrm{opt6}},\frac {|h_{sr}|^2}{|h_{rd}|^2}P_{sR}^{\textrm{opt6}}) \geq \widetilde f( \frac{|h_{rd}|^2}{|h_{sr}|^2} P_{rR}^{\textrm{opt8}},P_{rR}^{\textrm{opt8}})$&\\
\hline
$\neq 0$&$0$&0&$-$&$\widetilde f(P_{sR}^{\textrm{opt7}},\frac {|h_{sr}|^2}{|h_{rd}|^2}P_{sR}^{\textrm{opt7}}) \geq \widetilde f( \frac{|h_{rd}|^2}{|h_{sr}|^2} P_{rR}^{\textrm{opt5}},P_{rR}^{\textrm{opt5}})$&\multirow{4}{*}{$\left[ P_{sR}^{\textrm{opt7}},\frac {|h_{sr}|^2}{|h_{rd}|^2}P_{sR}^{\textrm{opt7}} \right]$}\\
\cline{1-5}
$\neq 0$&$0$&$\neq 0$&$>0$&$\widetilde f(P_{sR}^{\textrm{opt7}},\frac {|h_{sr}|^2}{|h_{rd}|^2}P_{sR}^{\textrm{opt7}}) \geq \widetilde f( \frac{|h_{rd}|^2}{|h_{sr}|^2} P_{rR}^{\textrm{opt6}},P_{rR}^{\textrm{opt6}})$&\\
\cline{1-5}
$\neq 0$&$0$&$\neq 0$&$0$&$\widetilde f(P_{sR}^{\textrm{opt7}},\frac {|h_{sr}|^2}{|h_{rd}|^2}P_{sR}^{\textrm{opt7}}) \geq \widetilde f( \frac{|h_{rd}|^2}{|h_{sr}|^2} P_{rR}^{\textrm{opt7}},P_{rR}^{\textrm{opt7}})$&\\
\cline{1-5}
$\neq 0$&$0$&$\neq 0$&$<0$&$\widetilde f(P_{sR}^{\textrm{opt7}},\frac {|h_{sr}|^2}{|h_{rd}|^2}P_{sR}^{\textrm{opt7}}) \geq \widetilde f( \frac{|h_{rd}|^2}{|h_{sr}|^2} P_{rR}^{\textrm{opt8}},P_{rR}^{\textrm{opt8}})$&\\
\hline
$\neq 0$&$<0$&0&$-$&$\widetilde f(P_{sR}^{\textrm{opt8}},\frac {|h_{sr}|^2}{|h_{rd}|^2}P_{sR}^{\textrm{opt8}}) \geq \widetilde f( \frac{|h_{rd}|^2}{|h_{sr}|^2} P_{rR}^{\textrm{opt5}},P_{rR}^{\textrm{opt5}})$&\multirow{4}{*}{$\left[ P_{sR}^{\textrm{opt8}},\frac {|h_{sr}|^2}{|h_{rd}|^2}P_{sR}^{\textrm{opt8}} \right]$}\\
\cline{1-5}
$\neq 0$&$<0$&$\neq 0$&$>0$&$\widetilde f(P_{sR}^{\textrm{opt8}},\frac {|h_{sr}|^2}{|h_{rd}|^2}P_{sR}^{\textrm{opt8}}) \geq \widetilde f( \frac{|h_{rd}|^2}{|h_{sr}|^2} P_{rR}^{\textrm{opt6}},P_{rR}^{\textrm{opt6}})$&\\
\cline{1-5}
$\neq 0$&$<0$&$\neq 0$&$0$&$\widetilde f(P_{sR}^{\textrm{opt8}},\frac {|h_{sr}|^2}{|h_{rd}|^2}P_{sR}^{\textrm{opt8}}) \geq \widetilde f( \frac{|h_{rd}|^2}{|h_{sr}|^2} P_{rR}^{\textrm{opt7}},P_{rR}^{\textrm{opt7}})$&\\
\cline{1-5}
$\neq 0$&$<0$&$\neq 0$&$<0$&$\widetilde f(P_{sR}^{\textrm{opt8}},\frac {|h_{sr}|^2}{|h_{rd}|^2}P_{sR}^{\textrm{opt8}}) \geq \widetilde f( \frac{|h_{rd}|^2}{|h_{sr}|^2} P_{rR}^{\textrm{opt8}},P_{rR}^{\textrm{opt8}})$&\\
\hline
0&$-$&0&$-$&$\widetilde f(P_{sR}^{\textrm{opt5}},\frac {|h_{sr}|^2}{|h_{rd}|^2}P_{sR}^{\textrm{opt5}}) \leq \widetilde f( \frac{|h_{rd}|^2}{|h_{sr}|^2} P_{rR}^{\textrm{opt5}},P_{rR}^{\textrm{opt5}})$&\multirow{4}{*}{$\left[ \frac{|h_{rd}|^2}{|h_{sr}|^2} P_{rR}^{\textrm{opt5}},P_{rR}^{\textrm{opt5}} \right]$}\\
\cline{1-5}
$\neq 0$&$>0$&0&$-$&$\widetilde f(P_{sR}^{\textrm{opt6}},\frac {|h_{sr}|^2}{|h_{rd}|^2}P_{sR}^{\textrm{opt6}}) \leq \widetilde f( \frac{|h_{rd}|^2}{|h_{sr}|^2} P_{rR}^{\textrm{opt5}},P_{rR}^{\textrm{opt5}})$&\\
\cline{1-5}
$\neq 0$&$0$&0&$-$&$\widetilde f(P_{sR}^{\textrm{opt7}},\frac {|h_{sr}|^2}{|h_{rd}|^2}P_{sR}^{\textrm{opt7}}) \leq \widetilde f( \frac{|h_{rd}|^2}{|h_{sr}|^2} P_{rR}^{\textrm{opt5}},P_{rR}^{\textrm{opt5}})$&\\
\cline{1-5}
$\neq 0$&$<0$&0&$-$&$\widetilde f(P_{sR}^{\textrm{opt8}},\frac {|h_{sr}|^2}{|h_{rd}|^2}P_{sR}^{\textrm{opt8}}) \leq \widetilde f( \frac{|h_{rd}|^2}{|h_{sr}|^2} P_{rR}^{\textrm{opt5}},P_{rR}^{\textrm{opt5}})$&\\
\hline
0&$-$&$\neq 0$&$>0$&$f(P_{sR}^{\textrm{opt5}},\frac {|h_{sr}|^2}{|h_{rd}|^2}P_{sR}^{\textrm{opt5}}) \leq f( \frac{|h_{rd}|^2}{|h_{sr}|^2} P_{rR}^{\textrm{opt6}},P_{rR}^{\textrm{opt6}})$&\multirow{4}{*}{$\left[ \frac{|h_{rd}|^2}{|h_{sr}|^2} P_{rR}^{\textrm{opt6}},P_{rR}^{\textrm{opt6}} \right]$}\\
\cline{1-5}
$\neq 0$&$>0$&$\neq 0$&$>0$&$\widetilde f(P_{sR}^{\textrm{opt6}},\frac {|h_{sr}|^2}{|h_{rd}|^2}P_{sR}^{\textrm{opt6}}) \leq \widetilde f( \frac{|h_{rd}|^2}{|h_{sr}|^2} P_{rR}^{\textrm{opt6}},P_{rR}^{\textrm{opt6}})$&\\
\cline{1-5}
$\neq 0$&$0$&$\neq 0$&$>0$&$\widetilde f(P_{sR}^{\textrm{opt7}},\frac {|h_{sr}|^2}{|h_{rd}|^2}P_{sR}^{\textrm{opt7}}) \leq \widetilde f( \frac{|h_{rd}|^2}{|h_{sr}|^2} P_{rR}^{\textrm{opt6}},P_{rR}^{\textrm{opt6}})$&\\
\cline{1-5}
$\neq 0$&$<0$&$\neq 0$&$>0$&$\widetilde f(P_{sR}^{\textrm{opt8}},\frac {|h_{sr}|^2}{|h_{rd}|^2}P_{sR}^{\textrm{opt8}}) \leq \widetilde f( \frac{|h_{rd}|^2}{|h_{sr}|^2} P_{rR}^{\textrm{opt6}},P_{rR}^{\textrm{opt6}})$&\\
\hline
0&$-$&$\neq 0$&$0$&$\widetilde f(P_{sR}^{\textrm{opt5}},\frac {|h_{sr}|^2}{|h_{rd}|^2}P_{sR}^{\textrm{opt5}}) \leq \widetilde f( \frac{|h_{rd}|^2}{|h_{sr}|^2} P_{rR}^{\textrm{opt7}},P_{rR}^{\textrm{opt7}})$&\multirow{4}{*}{$\left[ \frac{|h_{rd}|^2}{|h_{sr}|^2} P_{rR}^{\textrm{opt7}},P_{rR}^{\textrm{opt7}} \right]$}\\
\cline{1-5}
$\neq 0$&$>0$&$\neq 0$&$0$&$\widetilde f(P_{sR}^{\textrm{opt6}},\frac {|h_{sr}|^2}{|h_{rd}|^2}P_{sR}^{\textrm{opt6}}) \leq \widetilde f( \frac{|h_{rd}|^2}{|h_{sr}|^2} P_{rR}^{\textrm{opt7}},P_{rR}^{\textrm{opt7}})$&\\
\cline{1-5}
$\neq 0$&$0$&$\neq 0$&$0$&$\widetilde f(P_{sR}^{\textrm{opt7}},\frac {|h_{sr}|^2}{|h_{rd}|^2}P_{sR}^{\textrm{opt7}}) \leq \widetilde f( \frac{|h_{rd}|^2}{|h_{sr}|^2} P_{rR}^{\textrm{opt7}},P_{rR}^{\textrm{opt7}})$&\\
\cline{1-5}
$\neq 0$&$<0$&$\neq 0$&$0$&$\widetilde f(P_{sR}^{\textrm{opt8}},\frac {|h_{sr}|^2}{|h_{rd}|^2}P_{sR}^{\textrm{opt8}}) \leq \widetilde f( \frac{|h_{rd}|^2}{|h_{sr}|^2} P_{rR}^{\textrm{opt7}},P_{rR}^{\textrm{opt7}})$&\\
\hline
0&$-$&$\neq 0$&$<0$&$\widetilde f(P_{sR}^{\textrm{opt5}},\frac {|h_{sr}|^2}{|h_{rd}|^2}P_{sR}^{\textrm{opt5}}) \leq \widetilde f( \frac{|h_{rd}|^2}{|h_{sr}|^2} P_{rR}^{\textrm{opt8}},P_{rR}^{\textrm{opt8}})$&\multirow{4}{*}{$\left[ \frac{|h_{rd}|^2}{|h_{sr}|^2} P_{rR}^{\textrm{opt8}},P_{rR}^{\textrm{opt8}} \right]$}\\
\cline{1-5}
$\neq 0$&$>0$&$\neq 0$&$<0$&$\widetilde f(P_{sR}^{\textrm{opt6}},\frac {|h_{sr}|^2}{|h_{rd}|^2}P_{sR}^{\textrm{opt6}}) \leq \widetilde f( \frac{|h_{rd}|^2}{|h_{sr}|^2} P_{rR}^{\textrm{opt8}},P_{rR}^{\textrm{opt8}})$&\\
\cline{1-5}
$\neq 0$&$0$&$\neq 0$&$<0$&$\widetilde f(P_{sR}^{\textrm{opt7}},\frac {|h_{sr}|^2}{|h_{rd}|^2}P_{sR}^{\textrm{opt7}}) \leq \widetilde f( \frac{|h_{rd}|^2}{|h_{sr}|^2} P_{rR}^{\textrm{opt8}},P_{rR}^{\textrm{opt8}})$&\\
\cline{1-5}
$\neq 0$&$<0$&$\neq 0$&$<0$&$\widetilde f(P_{sR}^{\textrm{opt8}},\frac {|h_{sr}|^2}{|h_{rd}|^2}P_{sR}^{\textrm{opt8}}) \leq \widetilde f( \frac{|h_{rd}|^2}{|h_{sr}|^2} P_{rR}^{\textrm{opt8}},P_{rR}^{\textrm{opt8}})$&\\
\hline
\end{tabular}
\end{table*}

\section{Numerical Results and Discussions}
\indent This section presents the numerical secrecy rate results of our proposed JRJS schemes using FCSI-PA and PCSI-PA strategies. Pure relay selection, pure jamming, generalized singular-value-decomposition (GSVD) based beaforming and JRJS schemes using equal power allocation (EPA) are used as benchmark schemes. In our numerical experiments, we have $E(|h_{si}|^2)=E(|h_{id}|^2)=1$, $E(|h_{se}|^2)=\epsilon_1=1$, $E(|h_{ie}|^2)=\epsilon_2=1$ and $\delta_{n}^2=0$ dBm. We show that the proposed JRJS scheme outperforms the pure relay selection, pure jamming and GSVD based beamforming schemes. Also, our proposed FCSI-PA and PCSI-PA schemes both perform better than the EPA strategy in terms of secrecy rate. Moreover, numerical results illustrate that with an increasing number of intermediate nodes, the secrecy rates of the proposed JRJS schemes using FCSI-PA and PCSI-PA strategies increase.\\
\indent We first discuss the effect of $R_d$ on the secrecy rate of the proposed JRJS scheme. As stated in section $\textrm{\uppercase\expandafter{\romannumeral3}}$, transmission rate $R_d$ plays an important role in the secrecy rate results mainly due to two reasons. One reason is that different transmission rates lead to different decoding sets where the relay is selected, which can be seen from (10). The other reason is that secrecy rate results of the proposed FCSI-PA and PCSI-PA JRJS schemes are effected by $P_{b1}$ and $P_{b3}$ which both are the functions of $R_d$. It is noted that, in our experiments, we use $R_d \leq \log_2(\frac{E(|h_{sr}|^2)E(|h_{rd}|^2)}{E(|h_{sr}|^2)+E(|h_{rd}|^2)}\frac{P}{\delta_n^2}+1)$ instead of (26) as the guideline for $R_d$, which is easier to implement. Fig. 3 shows the secrecy rates of our proposed FCSI-PA and PCSI-PA JRJS schemes versus $R_d$ in the case of $P=14$ dBm. In Fig. 3, we can see that secrecy rate increases as $R_d$ increases from 1 bit/s/Hz to 3 bit/s/Hz and then decreases as $R_d$ increases from 3 bit/s/Hz to 4 bit/s/Hz.  The reason is that, when a higher transmission rate $R_d$ is used, $P_s|h_{sr}|^2$ needs to be larger to make sure that the relay can successfully decode the source signal, which will lead to a higher $C_d$ (see (11)) along with a higher secrecy rate. However, when $R_d$ is too high, $P_s$ needs to be even larger for the sake of successfully decoding the source signal, which would lead to a even smaller $P_r$ under the total power constraint. In this case, due to $C_{d}=\frac {1} {2} \log_2(1+\frac {\mathrm{min}(|h_{sr}|^2 P_s,|h_{rd}|^2 P_r)} {\delta_n^2})$ (see (11)), a lower $C_d$ along with a lower secrecy rate would be obtained. As shown in Fig. 3, the highest secrecy rate is obtained when $R_d=3$ bit/s/Hz is used in the case of $P=14$ dBm.\\
\indent Considering the effect of $R_d$ on secrecy rate results and the upper bound of $R_d$, we set $R_d=0.5$ bit/s/Hz, $R_d=1$ bit/s/Hz, $R_d=2$ bit/s/Hz, $R_d=3$ bit/s/Hz and $R_d=4$ bit/s/Hz in the case of $P \in \left[ 0, 3 \right]$ dBm, $P \in ( 3, 6]$ dBm, $P \in ( 6, 10]$ dBm, $P \in ( 10, 15]$ dBm and $P \in ( 15, 20]$ dBm, respectively, by experimental experience.\\
\begin{figure}[htb!]
\centering
\includegraphics[scale=0.65]{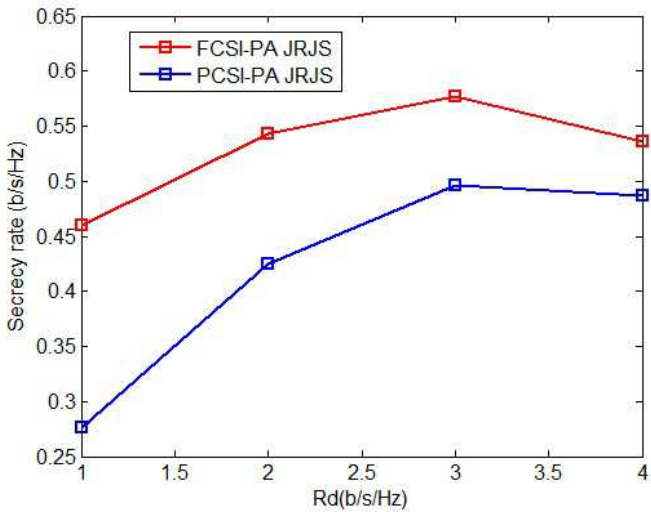}
\caption{Secrecy rate versus transmission rate $R_d$ of FCSI-PA JRJS and PCSI-PA JRJS in the case of $P=14$ dBm.}
\label{fig_3}
\end{figure}
\indent Fig. 4 and Fig. 5 show the secrecy rate results of our proposed JRJS schemes with FCSI-PA and PCSI-PA strategies, respectively,  versus the total power $P$ with $M=10$. Pure relay selection, pure jamming, direct transmission and GSVD based beaforming [37]-[38] schemes are used as benchmark schemes. The transmission rates used in the benchmark schemes are the same as those in the proposed JRJS schemes. In our experiments, we set $E(|h_{sd}|^2)=0.05$ so that the direct link from the source to the destination is negligible to match the wireless network illustrated in Fig. 1. In the pure relay selection scheme, the relay is selected in a similar way as that in the proposed JRJS scheme without considering jamming. Transmit power at the source and relay is equally allocated. In the pure jamming scheme, all the intermediate nodes are used as jammers to transmit null-steering artificial noise with the total power $\frac {P} {2}$ and the transmit power at the source is set to $\frac {P} {2}$. Since the relay selection is effected by $|h_{se}|^2$ in pure relay selection scheme, different relays may be selected under FCSI and PCSI assumptions, which are distinguished as FCSI and PCSI pure relay selection schemes, respectively. In the GSVD scheme, all the intermediate nodes in the decoding set are used as the relays for signal transmission with a total power $\frac {P} {2}$ and no artificial noise is used. In the GSVD scheme under the FCSI assumption (denoted as FCSI-GSVD), the transmission is performed based on GSVD of the instantaneous CSI of the channel from the relays to the destination and the channel from the relays to the eavesdropper, while in the GSVD scheme under the PCSI assumption (denoted as PCSI-GSVD), the transmission is performed based on GSVD of the instantaneous CSI of the channel from the relays to the destination and the statistical CSI of the channel from the relays to the eavesdropper. Here we compute $C_d$ in the GSVD scheme as the minimum of the channel capacity from the source to the relays and the channel capacity from the relays to the destination, i.e., $C_d=\mathrm{min}(C_{sr},C_{rd})$ (equation (7)). It is noted that when multiple intermediate nodes are selected as the relays, the channel capacity from the source to the relays should be the mininum of the channel capacity from the source to each relay so that all the relays can successfully decode the source signal and then transmit their re-encoded outcomes. \\
\indent As shown in Fig. 4 and Fig. 5, both the proposed FCSI-PA and PCSI-PA JRJS schemes outperform the pure relay selection, the pure jamming, direct transmission and the corresponding GSVD schemes in terms of secrecy rate, implying the security benefits of exploiting JRJS with power allocation to defend against eavesdropping attack. It is noted that under the FCSI assumption, the proposed JRJS scheme performs better than the GSVD scheme in terms of secrecy rate mainly due to that the former achieves a higher $C_d$. The reason is that the proposed FCSI-PA scheme selects a relay to forward the source signal, which can lead to a higher channel capacity from the source to the relay, compared to the FCSI-GSVD scheme which exploits multiple relays. Therefore, although the FCSI-GSVD scheme generally leads to a higher capacity from the relays to the destination when a lower $R_d$ is used, it still achieves a lower $C_d$ since $C_d=\mathrm{min}(C_{sr},C_{rd})$. \\
\begin{figure}[htb!]
\centering
\includegraphics[scale=0.65]{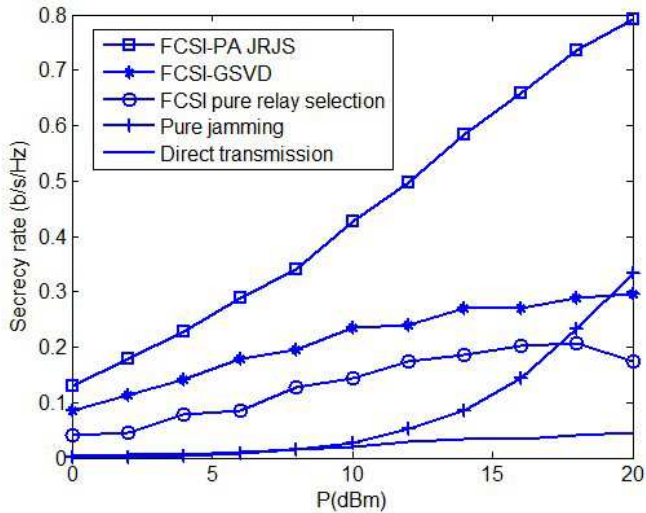}
\caption{Secrecy rate versus total transmit power $P$ of FCSI-PA JRJS, FCSI-GSVD based beamforming, FCSI pure relay selection, pure jamming and direct transmission schemes in the case of $M=10$.}
\label{fig_4}
\end{figure}
\begin{figure}[htb!]
\centering
\includegraphics[scale=0.65]{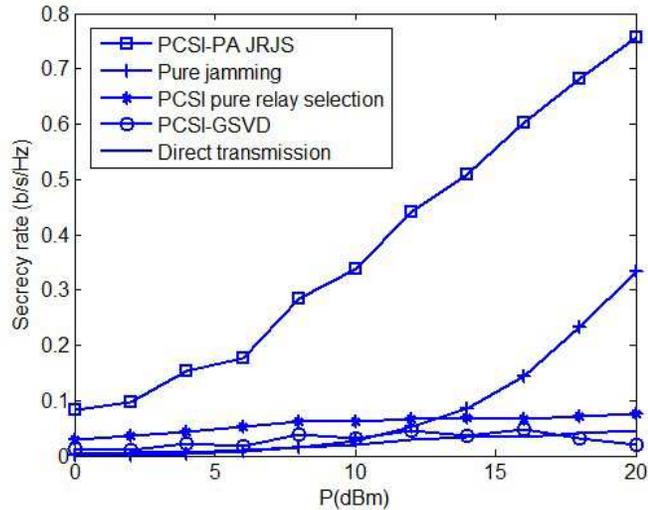}
\caption{Secrecy rate versus total transmit power $P$ of PCSI-PA JRJS, PCSI-GSVD based beamforming, PCSI pure relay selection, pure jamming and direct transmission schemes in the case of $M=10$.}
\label{fig_5}
\end{figure}
\indent Fig. 6 shows the secrecy rate results of the proposed JRJS schemes using FCSI-PA, PCSI-PA and EPA strategies versus total power $P$ with $M=10$. In the EPA scheme, transmit power at the source $P_s$, the relay $P_r$ and all the jammers $P_z$ is specified as $\frac{P}{2}, \frac{P}{4}, \frac{P}{4}$. That is, the transmit power in the first and second phase are allocated equally; the transmit power at the relay and the total transmit power at all the jammers in the second phase is allocated equally. We have also tested another fixed power allocation scheme where the transmit power at each intermediate node is equally allocated, that is, $P_s=\frac{P}{2}, P_r=\frac{P}{2M}, P_z=\frac{P}{2}-\frac{P}{2M}$, and obtained lower secrecy rate results than those obtained using the aforementioned EPA scheme. As observed in Fig. 6, the proposed power allocation strategies outperform the EPA strategies under both the FCSI and PCSI assumptions, showing the efficiency of proposed power allocation schemes. \\
\begin{figure}[htb!]
\centering
\includegraphics[scale=0.65]{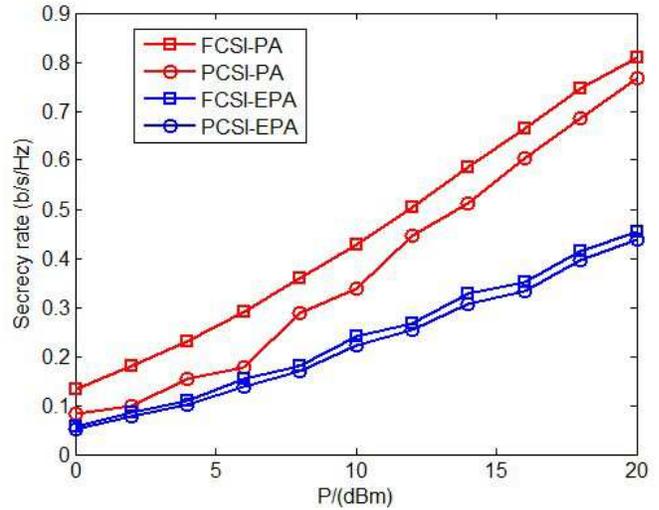}
\caption{Secrecy rate versus total transmit power $P$ of FCSI-PA, PCSI-PA, FCSI-EPA and PCSI-EPA JRJS schemes in the case of $M=10$.}
\label{fig_6}
\end{figure}
\indent Fig. 7 shows the relationship between the solved sub-optimal transmit power at source/relay and the number of intermediate nodes $M$ in the case of $P=14$ dBm. Power ratios $P_s/P$ and $P_r/P$ are used. As shown in Fig. 7, both $P_s/P$ and $P_r/P$ decrease as $M$ increases from 3 to 10 and keep almost unchanged when $M$ increases from 10 to 20. That means, for the sake of maximizing the secrecy rate of proposed JRJS scheme, the power allocated to transmit the source signal should increase with $M$ and tend towards almost fixed when $M>10$ given the total power $P$.\\
\begin{figure}[htb!]
\centering
\includegraphics[scale=0.65]{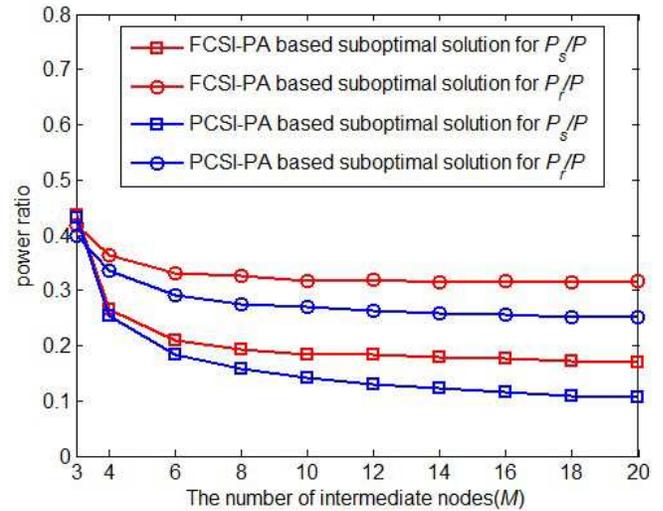}
\caption{sub-optimal power ratios $P_s/P$ and $P_r/P$ versus the number of intermediate nodes $M$ of FCSI-PA and PCSI-PA JRJS schemes in the case of $P=14$ dBm.}
\label{fig_7}
\end{figure}
\indent Fig. 8 shows the secrecy rate results of our proposed FCSI-PA and PCSI-PA JRJS schemes versus the number of intermediate nodes $M$ in the case of $P=14$ dBm. It can be easily seen that, the secrecy rates of proposed JRJS schemes improve notably as $M$ increases. It means that the physical-layer security of wireless communications relying on the proposed JRJS schemes can be further enhanced by employing more intermediate nodes. \\
\begin{figure}[htb!]
\centering
\includegraphics[scale=0.65]{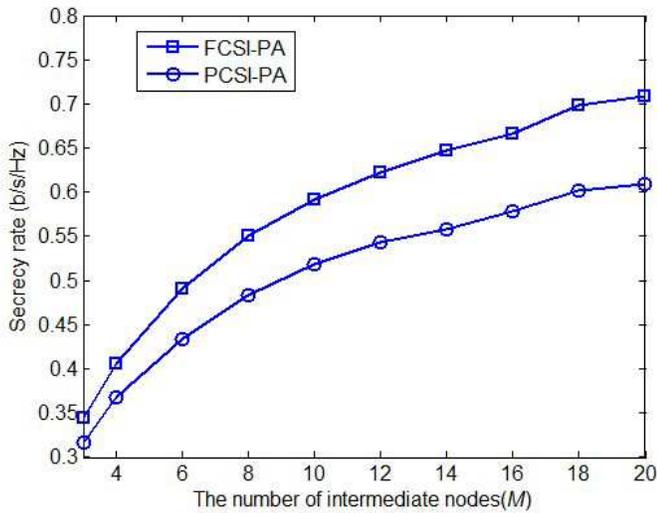}
\caption{Secrecy rate versus the number of intermediate nodes $M$ of FCSI-PA and PCSI-PA JRJS schemes in the case of $P=14$ dBm.}
\label{fig_8}
\end{figure}

\section{Conclusion}
\indent In this paper, we proposed a new JRJS scheme for improving the physical-layer security of a wireless DF relay systems. Given multiple intermediate nodes available, the proposed JRJS scheme selects one node to act as the relay, while the remaining intermediate nodes are enabled as the friendly jammers for transmitting artificial noise against the eavesdropper. We examined the power allocation among the source, relay and friendly jammers to maximize the secrecy rate of proposed JRJS scheme with a total power constraint. We derived closed-form sub-optimal solutions to the formulated power allocation problems under FCSI and PCSI assumptions, respectively. The relay and jammer selection as well as power allocation were considered jointly. Numerical results showed that the proposed JRJS framework outperforms the conventional pure relay selection, pure jamming and GSVD based beamforming methods in terms of secrecy rate. Also, the proposed FCSI-PA and PCSI-PA schemes are shown to achieve higher secrecy rates than the corresponding EPA strategies. Moreover, the secrecy rate of proposed JRJS framework relying on the FCSI-PA and PCSI-PA schemes can be improved by increasing the number of intermediate nodes.\\

\ifCLASSOPTIONcaptionsoff
  \newpage
\fi

\begin{IEEEbiography}[{\includegraphics[width=1in,height=1.25in]{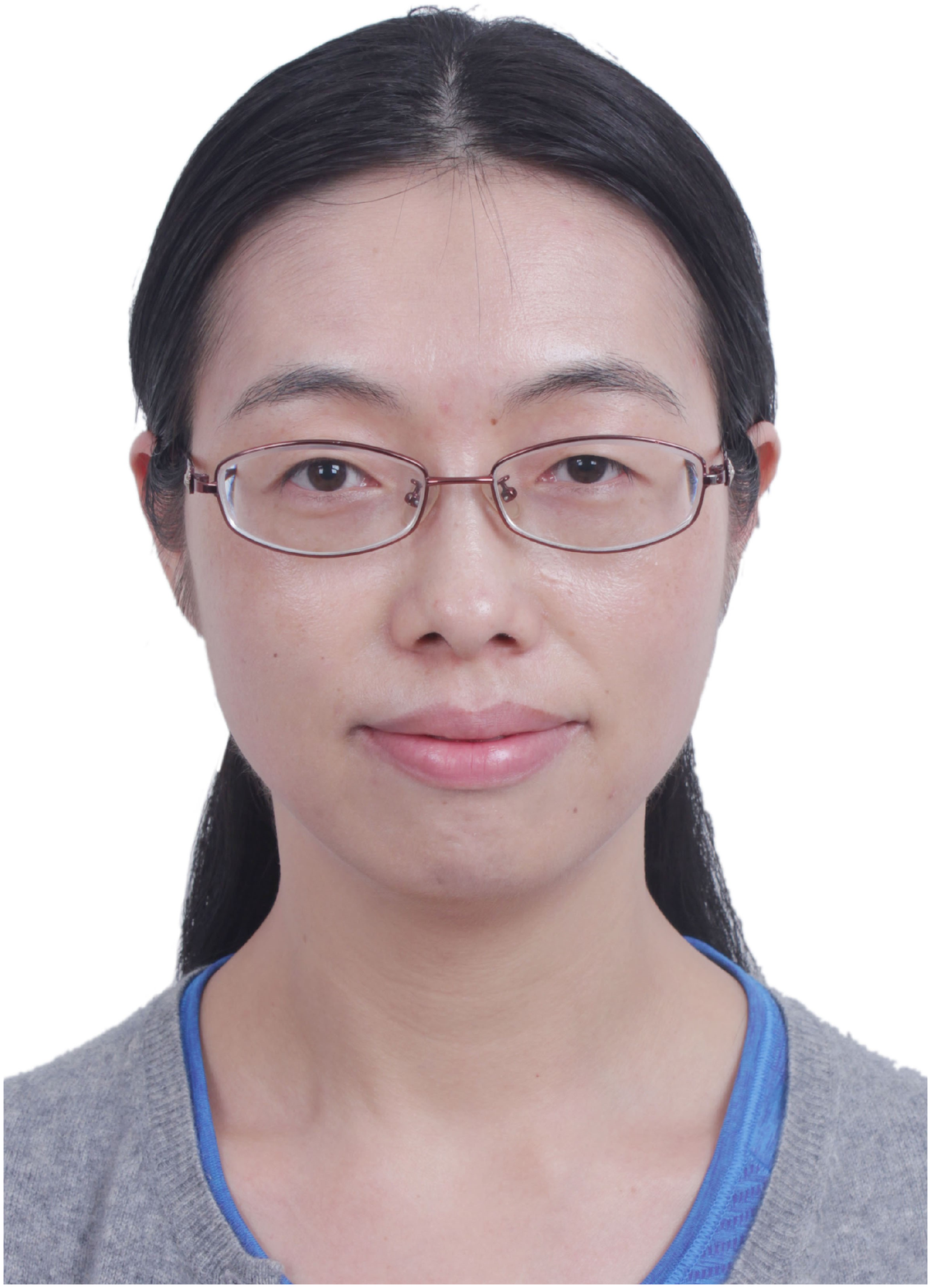}}]{Haiyan Guo} is a lecturer at the Nanjing University of Posts and Telecommunications (NUPT), Nanjing, China. She received her B.Eng. and Ph.D. degrees in signal and information processing from NUPT, Nanjing in 2005 and 2011, respectively. From 2013 to 2014, she was a post-doctoral research fellow with Southeast University. Her research interests include physical-layer security and speech signal processing.
\end{IEEEbiography}
\begin{IEEEbiography}[{\includegraphics[width=1in,height=1.25in]{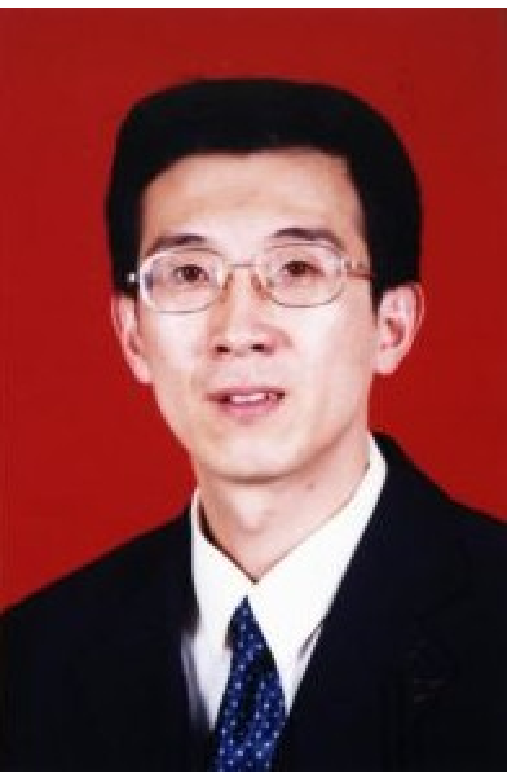}}]{Zhen Yang} is a professor at the Nanjing University of Posts and Telecommunications (NUPT), Nanjing, China. He received his B.Eng. and M.E. degrees in electrical engineering from NUPT, Nanjing, China, in 1983 and 1988, respectively, and the Ph.D. degree in electrical engineering from Shanghai Jiao Tong University, Shanghai, China, in 1999. From 1992 to 1993, he was a visiting scholar in Bremen University, Bremen, Germany, and in 2003, he was an exchange scholar in the University of Maryland, College Park.
His research interests include various aspects of signal processing and communication, such as communication systems and networks, cognitive radio, spectrum sensing, speech and audio processing, compressive sensing and wireless communication. Dr. Yang served as the Vice Chair of the Chinese Institute of Communications and was a member of the editorial board for several journals, including Chinese Journal of Electronics, Journal on Communications, and China Communications.
\end{IEEEbiography}
\vfill
\newpage
\begin{IEEEbiography}[{\includegraphics[width=1in,height=1.25in]{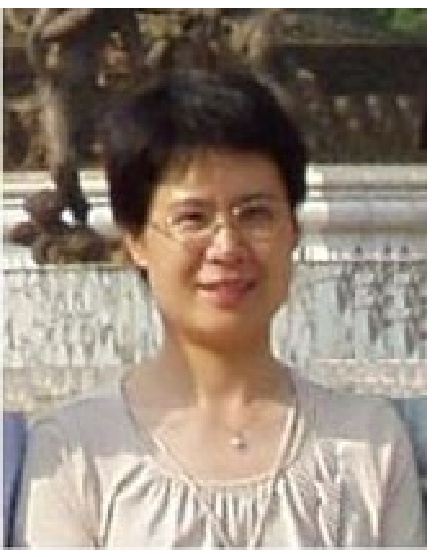}}]{Linghua Zhang} is a professor at the Nanjing University of Posts and Telecommunications (NUPT), Nanjing, China. She received her B.Eng. and M.E. degrees in Radio Engineering from Nanjing Engineering College (predecessor of Southeast University) in 1987 and 1990, respectively. She received her Ph.D. degree in signal and information processing from NUPT, Nanjing, China, in 2005. Her research interests include signal processing in wireless communication networks, modern speech communications and speech signal processing.
\end{IEEEbiography}
\begin{IEEEbiography}[{\includegraphics[width=1in,height=1.25in]{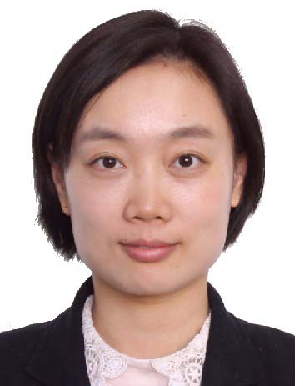}}]{Jia Zhu}  is a vice professor at the Nanjing University of Posts and Telecommunications (NUPT), Nanjing, China. She received the B.Eng. degree in computer science and technology from Hohai University, Nanjing, China, in 2005, and the Ph.D. degree in signal and information processing from NUPT, Nanjing, in 2010. From 2010 to 2012, she was a post-doctoral research fellow with the Stevens Institute of Technology, NJ, USA. Her general research interests include the cognitive radio, physical-layer security, and communications theory.
\end{IEEEbiography}
\begin{IEEEbiography}[{\includegraphics[width=1in,height=1.25in]{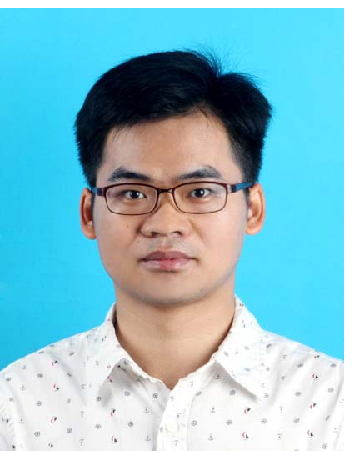}}]{Yulong Zou} (SM'13) is a professor at the Nanjing University of Posts and Telecommunications (NUPT), Nanjing, China. He received the B.Eng. degree in information engineering from NUPT, Nanjing, China, in July 2006, the first Ph.D. degree in electrical engineering from the Stevens Institute of Technology, New Jersey, USA, in May 2012, and the second Ph.D. degree in signal and information processing from NUPT, Nanjing, China, in July 2012.
His research interests span a wide range of topics in wireless communications and signal processing, including the cooperative communications, cognitive radio, wireless security, and energy-efficient communications. Dr. Zou was awarded the 9th IEEE Communications Society Asia-Pacific Best Young Researcher in 2014. He has served as an editor for the IEEE Communications Surveys \& Tutorials, IEEE Communications Letters, IET Communications, and China Communications. In addition, he has acted as TPC members for various IEEE sponsored conferences, e.g., IEEE ICC/GLOBECOM/WCNC/VTC/ICCC, etc.
\end{IEEEbiography}
\vfill


%\newpage


%\vfill



\begin{thebibliography}{1}

\bibitem{IEEEhowto:kopka}
Y. Zou, J. Zhu, X. Wang, and L. Hanzo, {``A survey on wireless security: Technical challenges, recent advances and future trends,''} \emph{Proc. IEEE,} vol.
104, no. 9, pp. 1727-1765, Sep. 2016.
\bibitem{IEEEhowto:kopka}
M. E. Hellman, {``An overview of public key cryptography,''} \emph{IEEE Commun. Mag.}, vol. 16, no. 6, pp. 42-49, May 2002.
\bibitem{IEEEhowto:kopka}
A. D. Wyner, {``The wire-tap channel,''} \emph{Bell Syst. Tech. J.}, vol. 54, no. 8, pp. 1355-1387, Oct. 1975.
\bibitem{IEEEhowto:kopka}
S. K. Leung-Yan-Cheong and M. E. Hellman, {``The Gaussian wiretap channel,''} \emph{IEEE Trans. Inf. Theory}, vol. 24, no. 4, pp. 451-456, Jul. 1978.
\bibitem{IEEEhowto:kopka}
F. Oggier and B. Hassibi, {``The secrecy capacity of the MIMO wiretap channel,''} \emph{IEEE Trans. Inf. Theory,}, vol. 57, no. 8, pp. 4961-4972, Oct. 2007.
\bibitem{IEEEhowto:kopka}
A. Khisti and G. W. Wornell, {``Secure transmission with multiple antennas-Part II:
the MIMOME wiretap channel,''} \emph{IEEE Trans. Inf. Theory,} vol. 56, no. 11, pp. 5515-5532, Nov. 2010.
\bibitem{IEEEhowto:kopka}
S. Yan, N. Yang, R. Malaney, and J. Yuan, {``Transmit antenna selection with alamouti coding and power allocation in MIMO wiretap channels ,''} \emph{IEEE Trans. Wireless Commun.,} vol. 13, no. 3, pp. 1656-1667, May 2014.
\bibitem{IEEEhowto:kopka}
N. Yang, S. Yan, J. Yuan, R. Malaney, R. Subramanian, and I. Land, {``Artificial noise: Transmission optimization in multi-input single-output wiretap channels,''} \emph{IEEE Trans. Commun.,} vol. 63, no. 5, pp. 1771-1783, May 2015.
\bibitem{IEEEhowto:kopka}
X. Chen, C. Zhong, C. Yuen, and H. Chen, {``Multi-antenna relay aided wireless physical layer security,''} \emph{IEEE Commun. Mag.,}, vol. 53, no. 16b, pp. 40-46, Dec. 2015.
\bibitem{IEEEhowto:kopka}
K. P. Peppas, N. C. Sagias and A. Maras, {``Physical layer security for multiple-antenna systems: a unified approach,''} \emph{IEEE Trans. Commun.,}, vol. 64, no. 1, pp. 314-328, Jan. 2016.
\bibitem{IEEEhowto:kopka}
Y. Zou, J. Zhu, X. Wang, and V. C.M. Leung, {``Improving physical-layer security in wireless communications using diversity techniques,''} \emph{IEEE Network,}, vol. 29, no. 1, pp. 42-48, 2015.
\bibitem{IEEEhowto:kopka}
L. Dong, Z. Han, A. P. Petropulu and H. V. Poor, {``Improving wireless physical layer security via cooperating relays,''} \emph{IEEE Trans. Signal Process.}, vol. 58, no. 3, pp. 1875-1888, Mar. 2010.     \bibitem{IEEEhowto:kopka}
J. Li, A. P. Petropulu and S. Weber, {``On cooperative relaying schemes for wireless physical layer security,''} \emph{IEEE Trans. Signal Process.}, vol. 59, no. 10, pp. 4985-4997, Oct. 2011.
\bibitem{IEEEhowto:kopka}
Y. Zou, X. Wang and W. Shen, {``Optimal relay selection for physical-layer security in cooperative wireless networks,''} \emph{IEEE J Sel Area Comm}, vol. 31, no. 10, pp. 2099-2111, Oct. 2013.
\bibitem{IEEEhowto:kopka}
S. ShahbazPanahi and M. Dong, {``Achievable rate region under joint distributed beamforming and power allocation for two-way relay networks,''} \emph{IEEE Trans. Wireless Commun.}, vol. 11, no. 11, pp. 4026-4037, Nov. 2012.
\bibitem{IEEEhowto:kopka}
Z. Ding, K. K. Leung, D. L. Goeckel, and D. Towsley, {``On the application of cooperative transmission to secrecy communications,''} \emph{IEEE J. Sel. Areas Commun.}, vol. 30, no. 2, pp. 359-368, Feb. 2012.
\bibitem{IEEEhowto:kopka}
V. Havary-Nassab, S. Shahbazpanahi, and A. Grami, {``Optimal distributed beamforming for two-way relay networks,''} \emph{IEEE Trans. Signal Process.}, vol. 58, no. 3, pp. 1238-1250, Mar. 2010.
\bibitem{IEEEhowto:kopka}
Y. Feng, Z. Yang, W.-P. Zhu, Q. Li, and B. Lv, {``Robust cooperative secure beamforming for simultaneous wireless information and power transfer in amplify-and-forward relay networks,''} \emph{IEEE Trans. Veh. Tech.}, vol. PP, no. 99, pp. 1-1, 2016.
\bibitem{IEEEhowto:kopka}
G. Zheng, L. Choo, and K. Wong, {``Optimal cooperative jamming to enhance physical layer security using relays,''} \emph{IEEE Trans. Signal Process.}, vol. 59, no. 3, pp. 1317-1322, Mar. 2011.
\bibitem{IEEEhowto:kopka}
M. Lin, J. Ge, and Y. Yang, {``An effective secure transmission scheme for AF relay networks with two-hop information leakage,''} \emph{IEEE Commun Lett.}, vol. 17, no. 8, pp. 1676-1679, Aug. 2013.     \bibitem{IEEEhowto:kopka}
Y. Liu, J. Li, and A. Petropulu, {``Destination assisted cooperative jamming for wireless physical layer security,''} \emph{IEEE Trans. Inf. Forensics Security.}, vol. 8, no. 4, pp. 682-694, Apr. 2013.
\bibitem{IEEEhowto:kopka}
Y. Su, L. Jiang, and C. He, {``Joint relay selection and power allocation for full-duplex DF co-operative networks with outdated CSI,''} \emph{IEEE Commun Lett.}, vol. 20, no. 3, pp. 510-513, Mar. 2016.
\bibitem{IEEEhowto:kopka}
X. Gong, H. Long, H. Yin, F. Dong, and B. Ren, {``Robust amplify-and-forward relay beamforming for security with mean square error constraint,''} \emph{IET Commun.}, vol. 9, no. 8, pp. 1081-1087, 2015.
\bibitem{IEEEhowto:kopka}
F. S. Al-Qahtani, C. Zhong, and H. M. Alnuweiri, {``Opportunistic relay selection for secrecy enhancement in cooperative networks,''} \emph{IEEE Trans. Commun.}, vol. 63, no. 5, pp. 1756-1770, May 2015.
\bibitem{IEEEhowto:kopka}
C.L. Wang, T.N. Cho, and K.J. Yang, {``On power allocation and relay selection for a two-way amplify-and-forward relaying system,''} \emph{IEEE Trans. Commun.}, vol. 61, no. 8, pp. 3146-3155, Aug. 2013.
\bibitem{IEEEhowto:kopka}
J. Chen, L. Song, Z. Han and B. Jiao, {``Joint relay and jammer selection for secure decode-and-forward two-way relay communications,''} \emph{IEEE Globecom}, pp. 5-9, Houston, Texas, USA, Dec. 2011.     \bibitem{IEEEhowto:kopka}
J. Chen, R. Zhang, L. Song, Z. Han and B. Jiao, {``Joint relay and jammer selection for secure two-way relay networks,''} \emph{IEEE T Inf Foren Sec}, vol. 7, no. 1, pp. 310-320, Feb. 2012.
\bibitem{IEEEhowto:kopka}
H. Wang, M. Luo, X. Xia and Q. Yin, {``Joint cooperative beamforming and jamming to secure AF relay systems with individual power constraint and no eavesdropper's CSI,''} \emph{IEEE Signal Proc Let}, vol. 20, no. 1, pp. 39-42, Jan. 2013.
\bibitem{IEEEhowto:kopka}
H. Wang, Q. Yin, W. Wang and X. Xia, {``Joint null-space beamforming and jamming to secure AF relay systems with individual power constraint,''} in \emph{Proc. ICASSP}, pp. 2911-2914, Vancouver, Canada,  May. 2013.
\bibitem{IEEEhowto:kopka}
L. Wang, Y. Cai, Y. Zou, W. Yang and L. Hanzo, {``Joint relay and jammer selection improves the physical layer security in the face of CSI feedback delays,''} \emph{IEEE Trans. Veh. Technol.}, vol. 65, no. 8, pp. 6259-6274, Aug. 2016.
\bibitem{IEEEhowto:kopka}
C. Wang, H. Wang and X. Xia, {``Hybrid opportunistic relaying and jamming with power allocation for secure cooperative networks,''} \emph{IEEE Trans. Wireless Commun.}, vol. 14, no. 2, pp. 589-605, Feb. 2015.
\bibitem{IEEEhowto:kopka}
T. Zheng, H. Wang, F. Liu and M. H. Lee, {``Outage constrained secrecy throughput maximization for DF relay networks,''} \emph{IEEE Trans. Commun.,} vol. 63, no. 5, pp. 1741-1755, May 2015.
\bibitem{IEEEhowto:kopka}
S. Yan, N. Yang, G. Geraci, R. Malaney, and J. Yuan, {``Optimization of code rates in SISOME wiretap channels,''} \emph{IEEE bTrans. Wireless Commun.,} vol. 14, no. 11, pp. 6377-6388, May 2015.
\bibitem{IEEEhowto:kopka}
J. Huang and A. L. Swindlehurst, {``Cooperative jamming for secure communications in MIMO relay networks,''} \emph{IEEE Trans. Signal Process.,} vol. 59, no. 10, pp. 4871-4884, Oct. 2011.
\bibitem{IEEEhowto:kopka}
A. Chorti, S. M. Perlaza, Z. Han and H. V. Poor, {``Physical layer security in wireless networks with passive and active eavesdroppers,''} in \emph{Proc. IEEE Global Telecommun. Conf.,} pp. 4868-4873, Dec. 2012.
\bibitem{IEEEhowto:kopka}
T. Kwon, V. W.S. Wong and R. Schober, {``Secure MISO cognitive radio system with perfect and imperfect CSI,''} \emph{Proc. IEEE Global Telecommun. Conf.,} pp. 1236-1241, Dec. 2012.
\bibitem{IEEEhowto:kopka}
A. Khisti, G. W. Wornell, A. Wiesel and Y. Eldar, {``On the Gaussian MIMO wiretap channel,''} \emph{Proc. IEEE Int. Symp. Inf. Theory (ISIT),} pp. 2471-2475, Jun. 2007.
\bibitem{IEEEhowto:kopka}
A. Khisti and G. W. Wornell, {``Secure transmission with multiple antennas I: The MISOME wiretap channel,''} \emph{IEEE Trans. Inf. Theory,} vol. 56, no. 7, pp. 3088-3104, Jul. 2010.
\end{thebibliography}
\end{document}